\documentclass[12pt]{article}
\usepackage{amsmath,amssymb,amsthm,amsfonts}
\pdfoutput=1
\usepackage{geometry}
\usepackage{cite}
\usepackage{framed}
\usepackage{tcolorbox}
\usepackage{adjustbox}
\usepackage{graphicx}
\usepackage{epsfig}
\usepackage{cases}
\usepackage{mathtools}
\usepackage{caption}
\usepackage{subcaption}
\usepackage[pagebackref=false,pageanchor=false,
unicode,psdextra]{hyperref}
\usepackage{bookmark}
\usepackage[affil-it]{authblk}
\usepackage{array}
\hypersetup{colorlinks=true,linkcolor=blue,filecolor=magenta,urlcolor=blue,citecolor=red}
\usepackage[title]{appendix}
\numberwithin{equation}{section}
\usepackage{extarrows}
\usepackage[colorinlistoftodos]{todonotes}
\usepackage{xcolor}
\usepackage{hyperref}
\def\hri#1#2{\href{http://arxiv.org/abs/#1}{[arXiv:#1]#2}}
\def\hre#1#2{\href{http://arxiv.org/abs/#1/#2}{[arXiv:#1/#2]}}

\def\a{\alpha}
\def\b{\beta}

\def\d{\delta}

\def\z{\zeta}

\def\k{\kappa}
\def\l{\lambda}
\def\m{\mu}
\def\n{\nu}

\def\r{\rho}

\def\s{\sigma}

\def\ph{\phi}

\def\D{\Delta}

\newcommand{\nn}{\nonumber}

\def\be{\begin{equation}}
\def\ee{\end{equation}}
\def\bea{\begin{eqnarray}}
\def\eea{\end{eqnarray}}
\def\bg{\begin{align}}
\def\eg{\end{align}}
\def\bs{\begin{split}}
\def\es{\end{split}}

\def\td{\tilde}

\def\ie{{\it i.e. }}

\def\eg{{\it e.g. }}
\begin{document}
\begin{titlepage}
{\title{{Higher order curvature corrections and holographic renormalization group flow}}}
\vspace{.5cm}
\author{Ahmad Ghodsi\thanks{a-ghodsi@ferdowsi.um.ac.ir}}
\author{Malihe Siahvoshan\thanks{malihe.siahvoshan@mail.um.ac.ir}}
\vspace{.5cm}
\affil{ Department of Physics, Faculty of Science,\\ 
	Ferdowsi University of Mashhad,  Mashhad, Iran}
\renewcommand\Authands{ and }
\maketitle
\thispagestyle{empty}
\vspace{-12cm}
\begin{flushright}
\end{flushright}
\vspace{10cm}
\begin{abstract} 
We study the holographic renormalization group (RG) flow in the presence of higher-order curvature corrections to the $(d+1)$-dimensional Einstein-Hilbert (EH) action for an arbitrary interacting scalar matter field by using the superpotential approach. We find the critical points of the RG flow near the local minima and maxima of the potential and show the existence of the bounce solutions. In contrast to the EH gravity, regarding the values of couplings of the bulk theory, superpotential may have both upper and lower bounds. Moreover, the behavior of the RG flow controls by singular curves. This study may shed some light on how a c-function can exist in the presence of these corrections.  
\end{abstract}

\end{titlepage}
\tableofcontents
\section{Introduction}
The Wilsonian approach in renormalization of quantum field theories (QFTs) \cite{Wilson:1971bg, Wilson:1971dh}, leads to the concept of the renormalization group (RG) by integrating out the high energy degrees of freedom. The RG flow connects different scale-invariant QFTs in the space of all the parameters (mass and couplings) of a QFT with a UV cut-off. 

The gauge/gravity duality \cite{Maldacena:1997re} provides a useful tool for study of the $d$-dimensional QFTs through the corresponding $(d+1) $-dimensional gravity theories.
It is specifically important when the quantum field theory is in the strong coupling regime in which the perturbation theory does not work.
Because of this correspondence, the RG flow of the QFT on the boundary is related to the various geometries of the bulk gravity and, a holographic dimension plays the role of the RG scale in the dual QFT. Consequently, the RG flow between two fixed points of the QFT is supposed to be given by a domain wall solution on the gravity side which connects two $AdS$ geometries. Usually, the RG flow obtains by perturbing the theory at a UV fixed point by a relevant or marginal operator. Adding these operators leads to a line of fixed points or an IR fixed point.

The main issue is the duality between the RG (Callan-Symanzik) equations and equations of the motion of the bulk action. The reason is that the RG equations are first-order differential equations and the corresponding equations of motion in the Einstein-Hilbert (EH) gravity are from the second order.

There are several approaches to solve this problem. The first attempts in \cite{Freedman:1999gp, Akhmedov:1998vf} showed the equivalency of the supergravity equation of motion and the RG flow equation of the super Yang-Mills boundary theory. 
Afterward, by applying the Hamilton-Jacobi formalism \cite{deBoer:1999tgo} found first-order equations of the classical supergravity action with the same form of the RG equations of the dual field theory, with even the correct contribution of the conformal anomalies. 
An alternative method to lower the order of the differential equations is using an auxiliary function (superpotential) \cite{Ceresole:2007wx}.  It has been applied in \cite{Kiritsis:2012ma, Bourdier:2013axa, Kiritsis:2014kua}. 
Despite the naming in the holographic RG flow literature, this function does not relate to the supersymmetric theories necessarily \cite{Skenderis:1999mm}. 
Moreover, due to the monotonic behavior of the superpotential, at least in the EH gravity, this function can be used as a candidate for $c$-function in the holographic study of the $c$-theorem \cite{Freedman:1999gp, Girardello:1998pd}.

In other approaches, \cite{Heemskerk:2010hk} finds a relation between the holographic RG flow and the Wilsonian RG flow in the corresponding field theory by using the role of multi-trace operators. Paper \cite{Faulkner:2010jy} shows an equivalence between integrating out the high energy degrees of freedom of the field theory and integrating out on a range of the holographic coordinate in the gravity side.

Using the superpotential method, \cite{Gursoy:2007cb, Gursoy:2008za} show a new possible exotic behavior of the RG flow. It has been used for Einstein-Hilbert bulk gravity in \cite{Kiritsis:2016kog}.
In the superpotential formalism, the exotic RG flow results from the new possible zero points of the beta function (the bounce points).  
The sign of beta function changes as it crosses this point, and unlike the usual UV/IR fixed points of the theory, the value of the potential is not extremum here. In other words, the beta function has a branch cut in the bounce points.
 
The result of \cite{Kiritsis:2016kog} has been extended to the maximally symmetric solutions with non-zero curvature in \cite{Ghosh:2017big}.
Following these works, \cite{Gursoy:2018umf, Bea:2018whf} have considered the black hole solutions in Einstein-dilaton theory and studied the finite temperature exotic RG flows.
Also, the $ c $-theorem along the exotic RG flow has been studied in \cite{Park:2019pzo} for 3-dimensional Einstein-dilaton gravity by considering the entanglement entropy as a holographic $ c $-function. 
Among other issues, \cite{Cremonini:2020rdx} discusses the exotic behavior of the RG flow in the anisotropic geometries in Einstein-Maxwell-scalar theories or equivalently in the dual non-relativistic quantum field theories.

A notable suggestion about the holographic study of the RG flow is the consideration of the higher curvature corrections to the gravity side. Generally, the order of the equations of motion is more than two, so we may encounter the holographic RG equations that are not the first order either. The study of these equations and their solutions may help us to understand the RG equations of the dual QFTs at the strong coupling regime.

The higher-order curvature terms appear in various theories.  For example, in supergravity or string theory, the action contains a series of higher curvature terms. The first general non-trivial corrections are the quadratic curvature terms
\be\label{e1}
\mathcal{L}_{corr}=\a_1 R^2 + \a_2 R_{\m\n}^2 + \a_3 R_{\m\n\r\s}^2\,.
\ee
In general, the coefficients may depend on the other fields of the theory. Moreover, in string theory, couplings are suppressed by string length scale, $l_s$. If this length is smaller than the curvature length, $l_s^2R\ll 1$, then we are in the perturbative regime and we can truncate the series expansion of the curvature corrections. Otherwise, all higher-order corrections would also be relevant. 

On the other hand, there are various constraints on the couplings of these corrections. For instance, in five dimensional Gauss-Bonnet (GB) gravity ($\a_1=\a_3=-\a_2/4=\l_{GB}$), it has been shown that to avoid the naked singularity, the GB coupling,  $\l_{GB}$, has to be limited to $\l_{GB}\leq\frac14$. 
Moreover, the unitarity of the dual boundary theory, demands that 
 $-\frac{7}{36}\leq \l_{GB} \leq \frac{9}{100}$, for example see 
\cite{Buchel:2009sk, Ghodsi:2018vhq}.

There are some evidence that the higher curvature terms affect the holographic bounds.  For instance, \cite{Brigante:2007nu, Brigante:2008gz } show that the viscosity bound is violated for positive values of the GB coupling.
They also observed that for $\l_{GB}\geq \frac{9}{100}$ there are metastable quasiparticles on the boundary CFT  that can travel faster than the speed of light and violate causality.
The existence of the ghost modes and tachyonic excitations are generic in these theories. To avoid causality violations, the higher-derivative couplings should be suppressed by appropriate powers of the Planck length $l_P$ or the string length scale $l_s$. At least for small values of the couplings, the ghost or tachyonic modes are integrated out, and their degrees of freedom are beyond the QFT cut-off.

In this paper, we consider constant coefficients for higher curvature terms and do not impose any restrictive condition. Although, we can always apply the above physical constraints, and our calculations would be trustful as far as we restrict ourselves in the allowed regions of the couplings of the theory.

The holographic RG flow of the general quadratic curvature (GQC) gravity in a simple toy model has been studied in \cite{Ghodsi:2019xrx}\footnote{For further properties of this theory of gravity for example see \cite{Ghodsi:2020qqb, Ghodsi:2015gna, Anastasiou:2021swo}}. 
It shows the existence of the a-theorem for even-dimensional theories by finding the Wess-Zumino action, which originates from the spontaneous breaking of the conformal symmetry, by using a radial cut-off near the AdS boundary.

In this paper, we will develop our understanding of the holographic RG flow by looking at the effects of the general quadratic curvature terms on the flow.
Similar to \cite{Kiritsis:2016kog}, we have considered a scalar matter field that minimally coupled to gravity with an arbitrary potential.
We define superpotential and write the equations of motion in terms of it. 
As the gauge-gravity correspondence suggests, the RG transformations of the boundary QFT are dual to a domain wall solution that connects the UV/IR fixed points. 
The scale factor of this geometry is related to the energy scale of the dual QFT, and therefore to the beta function of a marginal/relevant operator on the boundary. 
Finding the zeros of this beta function gives the critical points of the corresponding dual QFT. We will explain these steps in more detail in section 2.

Usually, the first choice to study the higher curvature corrections is the GB gravity, in which the order of derivatives of the equations of motion is the same as the EH action.
Although generally, we have three independent couplings ($a_1, a_2, a_3$) in action \eqref{e1}, the equations of motion only depend on two specific combinations of these couplings, say ($\k_1, \k_2$). 
The GB gravity stays on a specific simple class of the GQC gravity when $\k_1=0$.  
The space of two-derivative theories is more significant, and we will study this space in the first attempt. Then we extend our results for the general case when the equations of motion are the fourth-order derivatives.

This paper is organized as follows: In section 2 we provide a setup for GQC gravity. In section3, we present the details of the calculations of the holographic RG flow in two-derivative theories. We find the superpotential,  scalar field and, the scale factor near the possible critical points perturbatively.
We address the behavior of the holographic RG flow for GQC gravity in section 4. The last section is the summary and conclusions.

\section{The general setup} 
As was told in the introduction, we are going to study the effect of the quadratic curvature terms in the bulk gravity on the holographic RG flow.
Nearly all steps of the calculations are based on the techniques of paper \cite{Kiritsis:2016kog}. 
Let's consider a bulk action in a general $(d+1)$-dimensional space-time, that in addition to the Hilbert-Einstein action, it includes the following quadratic curvature terms with three independent couplings  
\be
\label{bulkaction}
S_{bulk}=\int d^{d+1} x\sqrt{-g}\Big(R +\mathit{a}_1 R^2 + \mathit{a}_2 R_{\m\n}^2 + \mathit{a}_3 (R^2 - 4 R_{\m\n}^2+R_{\m\n\r\s}^2)-\frac12(\partial_{\m}\phi)^2-V(\phi)\Big)\,.
\ee
Moreover, we couple a real scalar matter field with an arbitrary potential to gravity minimally. 
To study the holographic RG flow, we suppose that the potential for all the possible values of the scalar field is negative everywhere. 
It guarantees the existence of a dual conformal field theory at UV or IR fixed points of the RG flow.
In other words, at the gravity side, we suppose that the scalar field has various stationary points, specifically at the UV and IR fixed points. Consequently, at these points, the classical solution of the equations of motion is the $AdS_{d+1}$ space-time.

By gauge-gravity correspondence, the RG transformations of the boundary quantum field theories are dual to the diffeomorphisms of the bulk gravity. Therefore, it would be reasonable to consider a solution for equations of motion of the bulk fields that smoothly connects the AdS space-times at the UV and IR fixed points.  Therefore, we begin with the following metric, which in particular is a holographic representation of the RG flow between the two holographic dual boundary CFTs 
\be\label{anzats}
ds^2=e^{2A(r)}\big(-dt^2+\sum_{i=1}^{d-1} dx_idx^i\big)+dr^2\,, \qquad \phi=\phi(r)\,.
\ee
Here $r$ is the holographic coordinate, and we suppose that the boundary is a $d$-dimensional Minkowski space. 
In this description, the warp factor is related to the energy scale of the holographic dual quantum field theory
\be\label{logE}
\log\frac{E}{E_0}=A(r)\,.
\ee
If we consider the relevant operator $\mathcal{O}$ on the boundary CFT, which couples to the boundary value of the bulk scalar field, \ie 
\be
\mathcal{L}=\mathcal{L}_{CFT}+\phi(t,x^i)\mathcal{O}(t,x^i)\,,
\ee 
then it generates an RG flow, that goes away from the UV fixed point. 
The coupling of this operator \ie $g(E)=\phi(r,t,x^i)$ is running according to the RG flow equation and its corresponding beta function is given by
\be\label{betag}
\beta(g)=\frac{d g(E)}{d \log E}\,.
\ee
Therefore we can compute the beta function by knowing the scale factor, $A(r)$, so we need to find the solution of the bulk fields. 
For the  field configuration in  \eqref{anzats}, the equations of motion from the variation of the action with respect to metric are given by
\begin{subequations}
\begin{align}
\label{EOMTrr}
\frac12 \phi'(r)^2- V(\phi) &=-d\Big(-(d-1)A'(r)^2+{\k_2} A'(r)^4\nn \\
 &~~~+{\k_1}\big(2 d  A'(r)^2 A''(r)- A''(r)^2+2  A^{(3)}(r) A'(r) \big)\Big)\,, \\ 
\frac12 \phi'(r)^2+ V(\phi) &=-d(d-1)A'(r)^2-2 (d-1) A''(r) \nn \\
&~~~+{\k_2}\big(d  A'(r)^4+4 A'(r)^2 A''(r)\big)+{\k_1}\big(2 A^{(4)}(r) \nn \\ 
&~~~+3 d  A''(r)^2+4 d A^{(3)}(r) A'(r)+ 2d^2A'(r)^2 A''(r) \big) \,,\label{EOMTtt}
\end{align}
\end{subequations}
where  the above equations just depend on two combinations of the couplings of the quadratic curvature terms \ie 
\be\label{k1k2}
\k_1=4d a_1+(d+1) a_2\,,\qquad 
\k_2=(d-3)\Big((d-1) (d-2) a_3 +d  ((d+1) a_1+a_2)\Big)\,.
\ee
According to the above results, we divide our problem into two different sections. In section 3, we suppose $\k_1=0$, then the equations of motion are at most the second derivatives of the fields with respect to $r$. Although the differential equations are in the same order as the Einstein-Hilbert equations of motion, we experience new interesting situations. In section 4, when $\k_1\neq 0$ the equations of motion have fourth-order derivatives of the field. It makes things a little complicated. However, the analysis of section 3 reveals more information and details for the latter case.
\section{Holographic RG flow: \texorpdfstring{$\kappa_1=0$}{\unichar{"3BA}\unichar{"2081} = 0} theories}
In this section, we consider a simplified version of the GQC and suppose that $\k_1=0$. In particular, it contains the well known $(d+1)$-dimensional Gauss-Bonnet gravity when $a_1=a_2=0$. 
Moreover, this theory in its general form includes the critical and $f(R)$ theories of gravity up to the quadratic curvature terms. 

Following the Einstein-Hilbert case \cite{Kiritsis:2016kog}, we define a  superpotential that makes the equations of motion as a set of first-order differential equations\footnote{Everywhere in this paper $\ph'=\frac{d\ph(r)}{dr}$ and $W'=\frac{dW(\ph)}{d\phi}$.}
\be
\label{SP}
W(\phi(r))=-2(d-1)A'(r)\,.
\ee
If we rewrite the equations of motion \eqref{EOMTrr} and \eqref{EOMTtt} in terms of the superpotential and subtract/add these equations, then the set of the second order differential equations will reduce to the following first order equations 
\begin{subequations}
\begin{align}
\label{PhiEq}
\phi'&=\Big(1-\frac{\k_2}{2(d-1)^3}W^2\Big)W'\,,\\
\label{SPE}
V(\phi)&=-\frac{d}{4(d-1)}\Big(1-\frac{\k_2}{4(d-1)^3}W^2\Big)W^2+\frac{1}{2}\Big(1-\frac{\k_2}{2(d-1)^3}W^2\Big)^2 {W'}^2\,.
\end{align}
\end{subequations}
The holographic RG flow or the beta function for coupling $\phi$ of the relevant operator $\mathcal{O}$ is obtained from equations \eqref{logE} and \eqref{betag}  via
\be
\label{BetaFunc}
\beta(\phi) = \frac{d\phi(r)}{dA(r)}=-2(d-1)\Big(1-\frac{\k_2}{2(d-1)^3}W^2\Big)\frac{W'}{W}\,,
\ee
where in the last equality we have substituted the equations \eqref{SP} and \eqref{PhiEq}.
This equation shows that, by restricting to the region where $\k_2 < 0$, the sign of beta function would only depend on the sign of $\frac{W'}{W}$. Furthermore, the monotonic behavior of the superpotential along the RG flow would be guaranteed. This can be seen by 
\be
\label{RG flow}
\frac{dW(\phi(r))}{dr}=\Big(1-\frac{\k_2}{2(d-1)^3}W^2\Big){W'}^2\,.
\ee
Based on the above relation, for a  negative coupling $\k_2$, we can define a monotonically decreasing holographic c-function constructed out of the superpotential. 
For example in \cite{Ghodsi:2019xrx}, for a specific constant $\a$, it is proportional to $\frac{1}{W^{d+1}}( W^2+\a W^4)$. However, for positive values of $\k_2$, one needs to consider an upper bound either for $W$ or $\k_2$, otherwise, this superpotential would not be a monotonic function.
Before proceeding further, we should emphasize two important issues:

1. There is a constraint on the superpotential from the equation of motion \eqref{SPE}. If we write it as
\be
\label{Wprim}
W'^2 = \frac{(d-1)^2}{2(\k_2 W^2 - 2 (d-1)^3)^2} \Big( 16 (d-1)^4 V  + 4 d (d-1)^3 W^2 - d \k_2 W^4 \Big)\,,
\ee 
then the right hand side must be positive which it means that regarding the sign of $\k_2$ there are upper or lower bounds on the value of the superpotential. Since we have supposed $V(\phi)<0$ then
\begin{subequations}
\begin{align}
\label{nb}
& W\ge B_{-}>0\,,\qquad \qquad\qquad \,\,\, \k_2<0\,, \\
\label{pb}
& B_{+}\ge W\ge B_{-}>0\,,\qquad 0<\k_2<-\frac{d(d-1)^2}{4V(\phi)}\,, 
\end{align}
\end{subequations}
where for simplicity, we have assumed a positive superpotential for all values of the scalar field\footnote{In general, equations of motion are invariant under $W\leftrightarrow -W$.}. The upper and lower bounds are defined as
\be
\label{bpm}
 B_{\pm}^2=\frac{2 (d-1)^3}{\k_2} \Big( 1 \pm \sqrt{1 + \frac{4\k_2}{d(d-1)^2} V(\ph) } \Big)\,.
\ee
2. The scalar curvature can be written in term of the superpotential
\begin{align}\label{curv}
\mathcal{R} &=
\frac{d}{d-1} \Big( \phi' W' -\frac{d+1}{4(d-1)} W^2  \Big)\nn \\
&=\frac{d}{d-1} \Big( \big( 1 - \frac{\k_2}{2(d-1)^3} W^2 \big) {W'}^2 -\frac{d+1}{4(d-1)} W^2  \Big)\,.
\end{align}
The last equality shows that we should have a finite value for $W'(\phi(r))$ to have a regular geometry or nonsingular curvature.
\subsection{Critical points for \texorpdfstring{$\kappa_2<0$}{\unichar{"3BA}\unichar{"2082} < 0}}\label{31}
The zero points of the beta function \eqref{BetaFunc} for $\k_2 < 0$ are equivalent to the points where $W'$ vanishes. To find a relation between these extremums of the superpotential and the scalar potential $V$, we can differentiate \eqref{SPE} with respect to the  $\ph$
\be
\label{SPE Derivative}
V'(\ph) = \big( 1 - \frac{\k_2 W^2}{2(d-1)^3}  \big) \Big( ( 1 - \frac{\k_2 W^2}{2(d-1)^3}  ) W'' - \frac{d}{2(d-1)} (1 + \frac{2\k_2 W'^2 }{d(d-1)^2} ) W \Big)W'\,.
\ee 
In an extremum of the superpotential, if $W''$ is finite, then $V'=0$. When $W''$ diverges, the potential is not necessarily an extremum. 
In the following subsections, we will analyze both of these cases separately, \ie

$\bullet$ $W'=V'=0$ and $W''$ is finite.

$\bullet$ $W'=0$ and $W''$ diverges but $V'$ is finite.

We should emphasize that there is a lower bound on the superpotential everywhere, as we showed in \eqref{nb} for negative values of $\k_2$.
\subsubsection{Local maxima of the potential}\label{311}
\begin{figure}[t]
\centering
\begin{subfigure}{0.5\textwidth}
\includegraphics[width=1\textwidth]{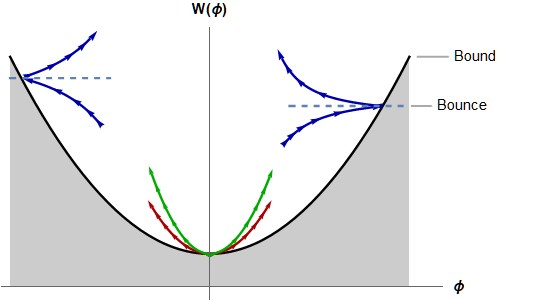}
\end{subfigure}
\caption{The generic behavior of the superpotential near the local maximum of the potential. The green curve is $W_+$ and the red one is for $W_-$. The blue curves represent the bounce solutions. The gray region is the forbidden area and is limited by the bound equation.}
\label{f1}
\end{figure}

Near a local maximum, the potential can be expanded as a sum of a cosmological term, a mass term ($m^2>0$), and interacting terms
\be
\label{Extremum V}
V(\phi)=-\frac{d(d-1)}{L^2} - \frac12 m^2 \phi^2 + \mathcal{O} (\phi^3)\,.
\ee 
For simplicity in the notation we have considered the extremum  at $\phi = 0 $. After substitution of the potential \eqref{Extremum V} in equation of motion \eqref{SPE}, two independent solutions appear for superpotential
\be
\label{WsolMax}
W_{\pm}(\phi)=\frac{2(d-1)}{\tilde{L}}+\frac{\Delta_{\pm}}{2\tilde{L}}\phi^2+ \mathcal{C}_{\pm} \phi^{z_{\pm}}  +\cdots\,,
\ee 
where dots denote the sub-leading terms and $\mathcal{C}_{\pm}$ are the constants of integration. 
The other parameters are defined as follow
\begin{subequations}
\label{newparameters}
\begin{align}
\tilde{L} &=\frac{L}{\sqrt{2}} \Big(1+\sqrt{1-\frac{4\k_2}{(d-1)L^2}}\Big)^{\frac12}\,,\\ 
\Delta_{\pm} &=\frac{d\pm\sqrt{d^2-4\tilde{L}^2 m^2}}{2\l}\,,\quad
\l=1-\frac{2\k_2}{(d-1)\tilde{L}^2}\,.
\label{newparameters2}
\end{align}
\end{subequations}
The reality condition for $\Delta_{\pm}$ implies that the mass is bounded from above, $0< m^2 \leq \frac{d^2}{4 \td{L}^2}$. Also $\k_2<0$ restricts $\l>1$ and therefore $L<\tilde{L}$.
As we will show, the new parameter $\tilde{L}$ is the radius of $AdS_{d+1}$ solution at the fixed point. For negative values of $\k_2$, it is larger than the radius of the $AdS_{d+1}$ solution of the Einstein-Hilbert action.
The power of the first leading terms are fixed by equations of motion
\be
z_{\pm}=\frac{d}{\l \Delta_{\pm}}\,, \qquad 0<z_{+}<2\,, \qquad 2<z_-\,,
\ee 
which consequently implies the boundary condition $\mathcal{C}_+=0$, otherwise $\phi=0$ would not be a fixed point because $W'$ diverges and the geometry is not smooth here.
In figure \ref{f1}, the generic behavior of the $W_{\pm}$ are depicted in green and red curves. These curves are limited to the region above the bound curve (gray region) \ie $W>B_-$.

As we already mentioned, the RG flow is given by \eqref{RG flow} and since $\frac{dW}{dr} \geq 0$ in this case, therefore $W$ and $r$ have the same increasing or decreasing behavior. The arrows in the figure \ref{f1}, show the increasing direction of the radial coordinate, or equivalently the RG flow direction. The outgoing RG flow from $\phi=0$ indicates that this critical point is a UV fixed point.

By solving the equation of motion \eqref{PhiEq} around $\ph = 0 $ and by using the superpotential solutions in \eqref{WsolMax}, the scalar field  solutions around the critical point can be found
\begin{subequations}
\begin{align}
&\phi_{+}(r)=\phi_{+} e^{\frac{\l \Delta_{+}}{\tilde{L}}r}+\cdots\,,\\
&\phi_{-}(r)=\phi_{-} e^{\frac{\l \Delta_{-}}{\tilde{L}}r}+
\frac{\mathcal{C}_- d \tilde{L}}{\Delta_{-}(d-2\l\Delta_{-})} \ph_{-}^{\Delta_+/\Delta_-}  e^{\frac{\l \Delta_{+}}{\tilde{L}}r} +\cdots\,,
\end{align}
\end{subequations}
where $\phi_{\pm}$ are the constants of the  integration. On the boundary field theory side we can associate two sources $J_\pm$ to the operator $\mathcal{O}$. In the standard quantization with using the above solution for scalar field, these sources are 
\be 
J_+=0\,,\qquad J_-=\frac{\phi_-}{\tilde{L}^{\l\Delta_-}}\,.
\ee
The vacuum expectation value of the operator $\mathcal{O}$ can be read as follow
\be  
\langle\mathcal{O}\rangle_+ =\frac{2\l\Delta_+ -d}{\tilde{L}^{\l\Delta_+}}\phi_+\,,\qquad 
\langle\mathcal{O}\rangle_- =\frac{d}{\tilde{L}^{\l\Delta_+}\Delta_-}\mathcal{C}\tilde{L}\phi_{-}^{\Delta_+/\Delta_-}\,.
\ee 
In the last step, we can read the geometry, near the fixed point, by using the definition of the superpotential \eqref{SP} 
\begin{subequations}
\begin{align}
A_{+}(r)=& -\frac{r-r_*}{\td{L}} - \frac{\phi_{+}^2}{8 \l (d-1)} e^{\frac{2\l \Delta_{+}}{\tilde{L}}r} +\cdots ,\\
A_{-}(r)=& -\frac{r-r_*}{\td{L}} - \frac{\phi_{-}^2}{8 \l (d-1)} e^{\frac{2\l \Delta_{-}}{\tilde{L}}r}+\mathcal{O}(e^{\frac{d r}{\td{L}}})+\cdots\,.
\end{align}
\end{subequations}
Since $\l\Delta_{\pm}>0$, both geometries are asymptotically $AdS$ spaces with radius $\tilde{L}$ when $r\rightarrow -\infty$. This is a UV fixed point and we expect a dual CFT.
\subsubsection{Local minima of potential}\label{312}
\begin{figure}[ht]
\centering
\begin{subfigure}{0.5\textwidth}
\includegraphics[width=1\textwidth]{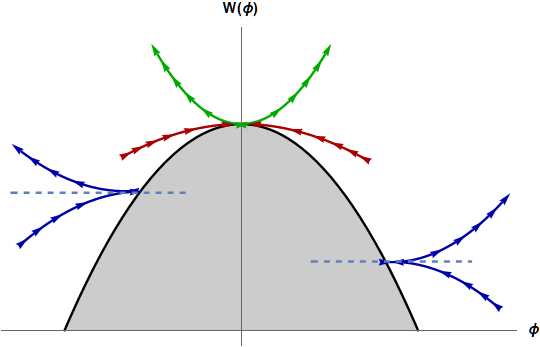}
\end{subfigure}
\caption{The generic behavior of the superpotential near the  local minimum of the potential. The green curve is $W_+$ and the red one is for $W_-$. The blue curves represent the bounce solutions.}
\label{f2}
\end{figure}

We can perform the same analysis as the previous section if the potential is locally near its minima ($m^2>0$)
\be
V(\phi)=-\frac{d(d-1)}{L^2} + \frac12 m^2 \phi^2 + \mathcal{O} (\phi^3),
\ee
We expect the same solutions for superpotential similar to the \eqref{WsolMax}, with an exception that $\Delta_{\pm}$ now are equal to
\be
\Delta_{\pm} = \frac{d\pm\sqrt{d^2+4\tilde{L}^2 m^2}}{2\l}.
\ee
Since we have considered $m^2>0$, the $\ph^z$ term is not the first leading term because
\be
z =z_\pm= \frac{d}{\l \Delta_\pm}\,,\qquad 0<z_+<1\,, \qquad z_-<0\,,
\ee 
therefore we should choose $\mathcal{C}_\pm = 0$. Finally, the solutions for scalar field and scale factor are given by
\begin{subequations} 
\begin{align}
\phi_{\pm}(r) &= \phi_{\pm} e^{\frac{\l \Delta_{\pm}}{\tilde{L}}r}+\cdots\,,\\
A_{\pm}(r) &= -\frac{r-r_*}{\td{L}} - \frac{\phi_{\pm}^2}{8 \l (d-1)} e^{\frac{2\l \Delta_{\pm}}{\tilde{L}}r} +\cdots \,.
\end{align}
\end{subequations}
In figure \ref{f2}, we have sketched the generic behavior of the superpotential (again green and red curves for $W_{\pm}$). The arrows on the curves show the increasing direction of the holographic coordinate $r$. 
So $\ph=0$ is an IR fixed point for $W_-$ solution, and at the same time, it is a UV fixed point for the $W_+$ branch. 
Acctually, $\Delta_+$ is positive and $\Delta_-$ is negative always, so the UV fixed point sits at $r\rightarrow -\infty$ but the IR fixed point is located at $r\rightarrow +\infty$. 
In other words, the endpoint of an RG flow is the starting point of another RG flow (see the same behavior and its related discussions for Einstein-Hilbert action in \cite{Kiritsis:2016kog}).
On the boundary field theory side the associate sources and vacuum expectation values of the operator $\mathcal{O}$ are as follow
\be  
J_{\pm} = 0\,,\qquad
\langle\mathcal{O}\rangle_\pm =\frac{2\l\Delta_\pm -d}{\tilde{L}^{\l\Delta_\pm}}\phi_\pm\,.
\ee 
\subsubsection{Bounces}\label{NegativeBounce}\label{313}
As we already discussed, there is another possibility for a critical point. The initial conditions for derivatives of the superpotential and potential  at this point are given by
\be 
W'(\ph_B) = 0 ,\qquad V'(\ph_B) \neq 0 ,\qquad \text{Divergent }W''(\ph_B).
\ee
It means that near the so-called bounce points \cite{Kiritsis:2016kog}, the superpotential is still at the extremum, however for a general potential with $V' \neq 0$, equation \eqref{SPE Derivative} leads to a divergent $W''$.
By differentiating the equation of motion \eqref{SPE} with respect to $\ph$ and  imposing the above conditions of the bounce point we get
\be
\label{Bounce EOM}
V'(\ph_B) \simeq (1 - \frac{\k_2}{2(d-1)^3} W^2 )^2 W' W''\,.
\ee 
Consider the following superpotential for some arbitrary constants $W_B$ and $\mathcal{C}_z$
\be \label{WBA}
W = W_B + \mathcal{C}_z (\ph - \ph_B)^z + \dots\,.
\ee 
By definition (equation \eqref{Wprim}), the lower bound is a set of points where $W'=0$ for a general potential function. Therefore, all the critical points belong to this set, and the extremum points of the potential are just specific points on this curve. 
Therefore, $W_B$ is the value of the superpotential at the bounce point, which is an arbitrary point on the lower bound curve  \eqref{nb} 
\be \label{VA}
W_B = B_-(\ph_B) = \sqrt{\frac{2 (d-1)^3}{\k_2}} \Big( 1 - \sqrt{1 + \frac{4\k_2}{d(d-1)^2} V(\ph_B) } \Big)^\frac12\,.
\ee
Substituting the ansatz \eqref{WBA} into the \eqref{Bounce EOM} and keeping the leading terms,  
we find that $z=\frac{3}{2}$ and the constant $\mathcal{C}_z$ is equal to
\be
\begin{split}
    \mathcal{C}_z^2 = \frac{8}{9} \Big(1 - \frac{\k_2}{2(d-1)^3} B_-(\ph_B)^2 \Big)^{-2} V'(\ph_B)
    = \frac{8}{9} \Big( 1+ \frac{4 \k_2}{d(d-1)^2} V(\ph_B) \Big)^{-1}  V'(\ph_B)\,,
\end{split}
\ee 
where in the second equality we have used the equation \eqref{VA}. Finally, we  have two distinct  solutions for the superpotential near the bounce point
\be\label{bkm}
\begin{split}
    W_{\uparrow,\downarrow}(\ph) &= B_-(\ph_B) \pm  \frac{2}{3} \Big( 1 - \frac{\k_2}{2(d-1)^3} B_-(\ph_B)^2  \Big)^{-1} \sqrt{2V'(\ph_B)} (\ph - \ph_B)^{\frac{3}{2}} + \dots\,.
\end{split}
\ee
The corresponding (blue) curves for $\uparrow$ and $\downarrow$ solutions are depicted in figures \ref{f1} and \ref{f2}. It should be noted that for positive(negative) $V'$, $\ph$ is greater (smaller) than $\ph_B$ \ie the bounces are reaching the bounds from the right(left) hand side.
These superpotentials, then lead to the following solution for scalar field near the bounce point. From equation \eqref{PhiEq} and the conditions of the bounce point we know that the $r$-derivative of the $\ph$ must be zero (since $W'=0$ at this point), so we find 
\be
\begin{split}
\ph_{\uparrow,\downarrow} (r) &=  \ph_B + \frac{9}{16} \Big(1 - \frac{\k_2}{2(d-1)^3} B_-(\ph_B)^2 \Big)^2 \mathcal{C}_z^2 (r - r_B)^2 + \mathcal{O} (r - r_B)^3 \\
&= \ph_B + \frac{V'(\ph_B)}{2} (r - r_B)^2 + \mathcal{O} (r - r_B)^3 \,,
\end{split}
\ee 
where $r_B$ is the corresponding radius of the bounce point and up (down) solution corresponds to  $r>r_B\, (r<r_B)$. The second line above is simplified by using the value of $\mathcal{C}_z$ and shows that at least at the first non-trivial order of the equations, the value of $\k_2$ does not play a role in the scalar field solution. 

The next step would be the determination of the geometry at this point. Using the definition of the superpotential in \eqref{SP} while we have the functionality of the superpotential and the scalar field near the bounce point we get
\be \label{bgeo}
\begin{split}
A(r) = 
 A_B - \frac{B_-(\ph_B)}{2(d-1)} (r - r_B) - \frac{V'(\ph_B)^2}{24 (d-1)} \Big(1- \frac{\k_2 B_-(\ph_B)^2}{2(d-1)^3}  \Big)^{-1} (r - r_B)^4  + \dots \,,
\end{split}
\ee
which is the same solution for the both branches of $W_{\uparrow,\downarrow}$.
As $r\rightarrow r_B$, the above geometry  tends to a regular geometry. 
The behavior of the beta function  \eqref{BetaFunc} near the bounce point is given by
\be
\begin{split}
    \b_{\uparrow,\downarrow}(\ph) 
= \pm \frac{2(d-1)}{B_-(\ph_B)} \sqrt{V'(\ph_B)} \sqrt{\ph - \ph_B} + \mathcal{O} (\ph - \ph_B)^2\,.
\end{split}
\ee 
Although the beta function is zero, it is not a fixed point, and actually, the RG flow does not reverse its direction at the bounce point.
It should be emphasized that the geometry in \eqref{bgeo},  unlike the local maxima and minima points, is a $(d+1)$-dimensional asymptotically flat space.
\subsection{Critical points for \texorpdfstring{$\kappa_2>0$}{\unichar{"3BA}\unichar{"2082} > 0}}\label{32}
In this section, we consider $\k_2>0$ and repeat all the previous steps in $\k_2>0$.
The beta function in equation \eqref{BetaFunc} for  $\k_2>0$ now has two types of fixed points. In addition to the extremum points of the superpotential, there are specific values for the supoerpotential at 
\be \label{WEpm}
W_E = \pm\sqrt{\frac{2 (d-1)^3}{\k_2}}\,,
\ee
where the beta function vanishes.
We should be careful since the beta function would be zero at $W=W_E$ only if $W'$ gets a finite value.  Therefore we divide our analysis into two subsections when $W$ is equal to  $W_E$ or not.

Before we proceed, it is worth reminding that according to the relation \eqref{pb} the superpotential is limited between two regions \ie $B_+\ge W \ge B_-$, provided that the coupling $\k_2$ has an upper bound
\be
\label{PotentialBound}
0<\k_2 \leq -\frac{d(d-1)^2}{4 V_{Max}}\,,
\ee
where $V_{Max}$ is supposed to be a global maximum of the potential, \ie $V(\phi)\leq V_{Max}<0$.
\subsubsection{Critical points for \texorpdfstring{$W \neq W_E$}{W\unichar{"2260}W\unichar{"1D07}} }\label{321}
\begin{figure}[ht]
\centering
\begin{subfigure}{0.5\textwidth}
\includegraphics[width=1\textwidth]{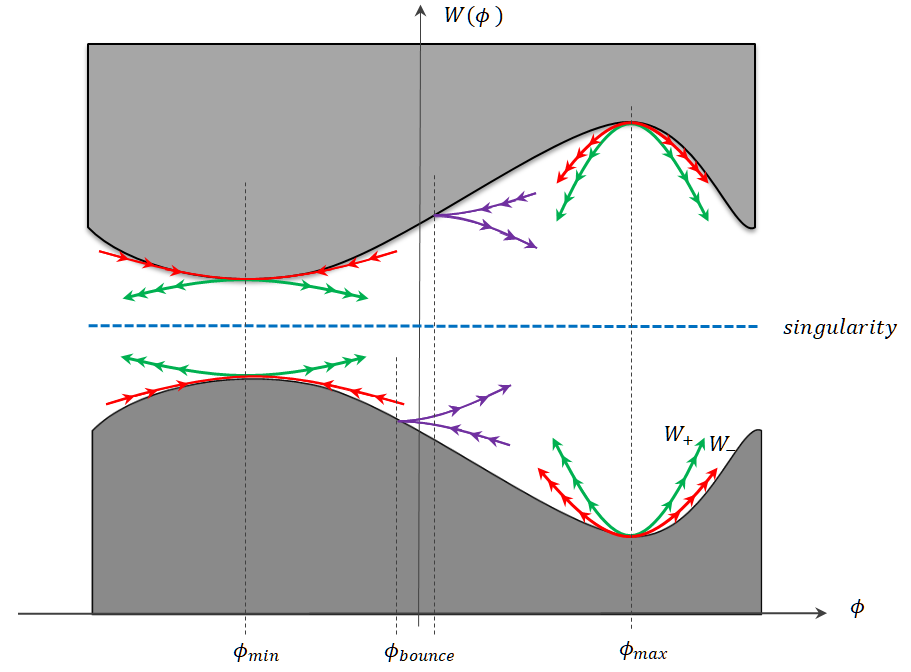}
\caption{$\,\,$}\label{f3}
\end{subfigure}
\centering
\begin{subfigure}{0.48\textwidth}
\includegraphics[width=1\textwidth]{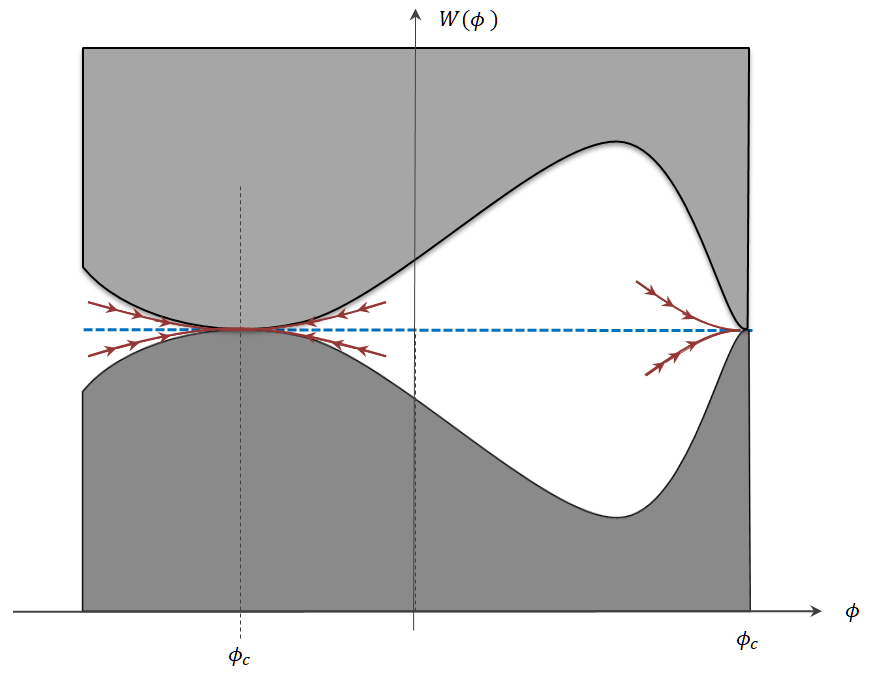}
\caption{$\,$}\label{f4}
\end{subfigure}
\caption{(a): The generic behavior of the superpotential near a local maximum $\ph_{max}$ or minimum $\ph_{min}$ of the potential and its RG flow are depicted as green and red curves. The local maxima are the UV fixed points, but the local minima are the UV fixed points for $W_-$ branch and IR fixed points for $W_+$. The bounce solutions also exist everywhere away from the extrema (purple curves). The blue dashed line is $W=W_E$, which is the location of curvature singularity. (b) In a specific point where the two bounds meet each other, the singularity removes, and an IR fixed point appears at the cross points.}
\end{figure}

Away from $W=W_E$, we have various critical points:  

$\bullet$ Local maxima of the potential: 
Near $\ph =\phi_{max}$ (see figure \ref{f3}), the solutions of the superpotential similarly are written as \eqref{WsolMax} with a slightly change in the parameters of the solutions. Unlike the negative coupling solutions, there are two acceptable value for $\td{L}$ for each branches. By choosing $\phi_{max}=0$ we have
\begin{subequations}
\begin{align}\label{Wu}
    W_{\pm}^{u}(\phi) &= \frac{2(d-1)}{\td{L}_{u}} + \frac{\Delta_{\pm}^{u}}{2\td{L}_{u}}\phi^2+ \mathcal{C}_{\pm}^{u} \phi^{u_\pm} + \cdots\,,\\
    W_{\pm}^{d}(\phi) &= \frac{2(d-1)}{\td{L}_{d}} + \frac{\Delta_{\pm}^{d}}{2\td{L}_{d}}\phi^2+ \mathcal{C}_{\pm}^{d} \phi^{v_\pm} + \cdots\,,
     \label{Wd}
\end{align}
\end{subequations}
where $\pm$ define the branches and $u$ and $d$ indices denote the solutions according to their starting points on the upper or lower bounds.
The parameters of the solutions are defined as follow
\begin{subequations}
\begin{align}
\label{Posnewparameters1}
&\td{L}_{(u,d)} = \frac{L}{\sqrt{2}} \Big( 1 \mp \sqrt{1 - \frac{4\k_2}{(d-1)L^2}} \Big)^\frac12 \,,\qquad 
\Delta_{\pm}^{(u,d)} = \frac{d\pm\sqrt{d^2-4\td{L}_{(u,d)}^2 m^2}}{2\l_{(u,d)}}\,,\\
& \l_{(u,d)} = 1-\frac{2\k_2}{(d-1) \td{L}_{(u,d)}^2}\,,\qquad u_\pm=\frac{d}{\Delta_{\pm}^u \l_u}\,,\qquad v_\pm=\frac{d}{\Delta_{\pm}^d \l_d}\,.
\label{Posnewparameters2}
\end{align}
\end{subequations}
The reality conditions of these parameters restrict the mass of the scalar field and the value of $\k_2$ coupling to 
\be
0<m\le \frac{d}{2\td{L}_{(u,d)}}\,, \qquad
0<\k_2 \le \frac{d-1}{4} L^2 \,.
\ee
Therefore the upper bound on $\k_2$ is determined via $\min\big(\frac{d-1}{4} L^2,-\frac{d(d-1)^2}{4 V_{Max}}\big)$.
The parameters of the solutions obey the following inequalities
\begin{subequations}
\begin{align}
\label{PosLTBound}
& 0 < \td{L}_u \leq \frac{L}{\sqrt{2}} \leq \td{L}_d < L\,,\\
&  
-1 \leq \l_u <0 \leq \l_d < 1\longrightarrow \Delta_\pm^u<0<\Delta_\pm^d\,, \\
 & 1<u_+<2<u_-\longrightarrow \mathcal{C}_+^u=\mathcal{C}_+^d=0\,. 
\end{align}
\end{subequations}
By using the above results, we have sketched the superpotential in figure \ref{f3}, in terms of the scalar field near $\ph=\ph_{max}$. In this figure, the gray regions are the forbidden areas of the superpotential due to the upper and lower bounds. The blue dashed line represents the critical value $W=W_E$. For those solutions above the $W_E$ line,  $\td{L}=\td{L}_u$  and for those under the line $\td{L}=\td{L}_d$. In figure \ref{f3}, all green curves are sketched for $W_+$ solution and the red ones for $W_-$. The analysis of RG flow shows that in $\phi=\phi_{max}$, there are two UV fixed points related to the up and down solutions of the superpotential, and the RG flow goes out of the local maxima points.

$\bullet$ Local minima of the potential:
Similar to the local maxima, there are two sets of solutions together with two values for $\td{L}$. 
We have the same solutions as \eqref{Wu} and \eqref{Wd} for $W>W_E$ and $W<W_E$ together with the same parameters as \eqref{Posnewparameters1} and \eqref{Posnewparameters2} but with a change  $m^2\rightarrow -m^2$.
The leading terms of the superpotential near $\phi=\phi_{min}=0$  is given by
\be 
W_{\pm}^{(u,d)}(\phi) = \frac{2(d-1)}{\tilde{L}_{(u,d)}} + \frac{\Delta_{\pm}^{(u,d)}}{2\tilde{L}_{(u,d)}}\phi^2 +\cdots \,.
\ee 
A generic behavior of the superpotential is depicted in the figure \ref{f3} near the point $\ph=\ph_{min}$. It should be noted that in this case,
\be 
\Delta_+^u<0<\Delta_-^u\,,\qquad \Delta_-^d<0<\Delta_+^d\,.
\ee
The critical point above the line of $W=W_E$ is an IR fixed point for $W_{-}^u$ branch (red curve) and a UV fixed point for $W_+^u$ branch (green curve). The same UV/IR behavior holds for $W_\pm^d$ below the line of $W=W_E$. 

$\bullet$ Bounces:
For positive values of $\k_2$, the bounces are sitting on both the upper and lower bounds. The solutions are
\be 
 W^{(u,d)}_{\uparrow,\downarrow}(\ph) = B_{\pm}(\ph_B) \pm  \frac{2}{3} \Big( 1 - \frac{\k_2}{2(d-1)^3} B_{\pm}(\ph_B)^2  \Big)^{-1} \sqrt{2V'(\ph_B)} (\ph - \ph_B)^{\frac{3}{2}} + \dots\,,
\ee 
where the upper and lower bounds, $B_\pm$, are given in equation \eqref{bpm}.
Two examples of the bounce solution are sketched, as the purple curves, in figure \ref{f3}.
\subsubsection{Critical points near \texorpdfstring{$W = W_E$}{W=W\unichar{"1D07}}}
\label{WequalWE}
As we discussed earlier, the beta function in \eqref{BetaFunc} or equivalently
\be 
\beta(\phi)=-2(d-1)\Big(1-\frac{W^2}{W_E^2}\Big)\frac{W'}{W}\,,
\ee
tends to zero at a specific critical value of the superpotential, $W_E=\sqrt{\frac{2 (d-1)^3}{\k_2}}$, if $W'$ has a finite value. On the other hand the equation of motion \eqref{SPE} can be written as follow
\be \label{SPEN}
V(\phi)=-\frac{d}{4(d-1)}\Big(1-\frac{1}{2}\frac{W^2}{W_E^2}\Big)W^2+\frac{1}{2}\Big(1-\frac{W^2}{W_E^2}\Big)^2 {W'}^2\,.
\ee 
Therefore for a finite $W'$ at $W=W_E$ we get
\be 
V(\ph_c) = V_c = - \frac{d(d-1)^2}{4\k_2}\,.
\ee 
If we insert $V_c$ into the equation of the bounds in \eqref{bpm}, then $W_E=B_+=B_-$. It means that this critical point exists only if the two bounds meet each other at $W=W_E$. 
However, if $V(\ph)\neq V_c$, then $W'$ tends to infinity, and therefore, we would have a curvature singularity. 
This singularity is shown as a blue dashed line in figures \ref{f3} and \ref{f4}. The RG flow never crosses the line of $W=W_E$, except at the specific points where the two bounds meet each other, see figure \ref{f4}.
As we see in \eqref{RG flow}, above and below the line of $W = W_E$, the RG flow changes its sign
\be \label{udwe}
W > W_E: \,\,\frac{dW}{dr} < 0\,,\qquad
W < W_E: \,\, \frac{dW}{dr} > 0\,,
\ee
It means that the RG flow stops on the cross points of the bounds at $\phi=\phi_c$. If there is no crossing point, then the RG flow tends to the $W=W_E$ line asymptotically from above and below as $r$ goes to $\pm\infty$.
To find the solutions of superpotential near the crossing points, we consider the following ansatz for potential and superpotential
\begin{subequations}
\begin{align}
&V(\ph) = V_c +V_1 (\ph-\ph_c)+V_2 (\ph-\ph_c)^2+\cdots \,,\\
& W(\ph) = W_E  + \sum_{i}\mathcal{C}_i (\ph-\ph_c)^{z_i}\,.
\end{align} 
\end{subequations}
For a finite value of $W'(\phi_c)$, the solution exists if $V_1=0$ and $V_2>0$. It means that $\phi_c$ is a local minimum of the potential. The superpotential is
\begin{subequations}
\begin{align}
W &=W_E+\mathcal{C}_1 (\phi-\phi_c)+\mathcal{C}_2 (\phi-\phi_c)^2+\cdots\,,\\
\mathcal{C}_1 &=\pm W_E \Big(\frac{d}{8(d-1)}\big(-1+\sqrt{1+32\frac{V_2}{W_E^2}\frac{(d-1)^2}{d^2}}\big)\Big)^\frac12\,,  \\
\mathcal{C}_2 &=\frac{2(d-1)V_2}{d W_E-3\sqrt{32(d-1)^2V_2+d^2 W_E^2}}\,,
\end{align}
\end{subequations}
and the scalar field and scale factor are given by
\begin{subequations}
\begin{align}
&\phi_{\pm}(r)=\phi_c+\phi_\pm e^{-2\frac{\mathcal{C}_1^2}{W_E}r}\pm\frac{|\mathcal{C}_1| d W_E}{12(d-1)\mathcal{C}_1^2+d W_E^2}(\phi_\pm e^{-2\frac{\mathcal{C}_1^2}{W_E}r})^2+\cdots\,,\\
& A_{\pm}(r)=-\frac{W_E}{2(d-1)}(r-r_*)\pm\frac{W_E}{2|\mathcal{C}_1|}\phi_\pm e^{-2\frac{\mathcal{C}_1^2}{W_E}r}+\cdots\,.
\end{align}
\end{subequations}
The last equation obviously shows that $\ph_c$ is an IR fixed point as $r\rightarrow+\infty$ and the radius of $AdS$ space is given by $\tilde{L}_E=\frac{2(d-1)}{W_E}$. Moreover, by using the definition of the superpotential in \eqref{SP}, $A_+(A_-)$ belongs to the region above (below) the line of $W=W_E$ as we discussed in \eqref{udwe}. Depending on how the RG flow approaches $\phi=\phi_c$ from the left or right, there might be four solutions, two correspond to the up and two for the down points of $W=W_E$. These solutions are depicted by brown curves in figure \ref{f4}. 
\section{Holographic RG flow: General case}\label{4}
In this section, we consider the general quadratic curvature action in which both $\k_1$ and $\k_2$ are non-zero. Although in these theories usually, one may find the spin two ghost modes (for example see \cite{Boulware:1985wk, Gullu:2009vy, Ghodsi:2017iee}) and the holographic dual quantum field theories suffer the lack of unitarity, nevertheless we can consider the coefficient to be small (suppressed by including a Planck or string length scale). We hope that the study of the holographic RG flow would give interesting information on the RG equations of the dual QFTs at strong coupling regime. 

To find the critical points of this theory, we use the same definition of the superpotential in \eqref{SP}, although it does not reduce the differential equations to the first-order ones. By writing the equations of motion in terms of the superpotential, we encounter with the higher-order derivatives of $\phi(r)$ which we can eliminate by using the equation of motion of the scalar field 
\be \label{phiz}
\phi''(r)=V'(\phi(r))-d A'(r)\phi'(r)\,.
\ee 
This equation is coming either from the variation of the Lagrangian with respect to the $\phi(r)$ or equivalently by a simple combination of the equations of motion \eqref{EOMTrr} and \eqref{EOMTtt} and their derivatives. Doing this, we get the following set of equations
\begin{subequations}
\begin{align}\label{GW1}
V& - \frac{1}{16 (d-1)^4}\Big( d \k_2 W^4-4d (d-1)^3 W^2  +    4 (d-1)^2 \big(2 (d-1)^2 - d \k_1 {W'}^2\big) {\phi'}^2 \nn \\
&\qquad\qquad\qquad\,\,\, + 8  d (d-1)^2\k_1 W \big(V' W' + {\phi'}^2 W''\big)\Big)=0\,, \\
\label{GW2}
V &+\frac{1}{16 (d-1)^4}\Big(-d \k_2 W^4 + 4 (d-1) W^2 \big(d(d-1)^2 + 2 \k_2 W' \phi'\big)- 8d (d-1)^2  \k_1 W (V' W' \nn \\
&\qquad\qquad\qquad\,\,\, -{\phi'}^2 W'') + 
 4 (d-1)^2 \phi'\big(-d \k_1 {W'}^2 \phi'- 4 (d-1) W' (d-1 - \k_1 V'') \nn \\
 &\qquad\qquad\qquad\,\,\, +2 (d-1) \big((d-1) \phi' + 
6 \k_1 V'  W'' + 2 \k_1 {\phi'}^2 W^{(3)}\big)\big)\Big)=0\,.
\end{align}
\end{subequations}
To extract the value of $\ph'$ it is enough to  subtract the above equations 
\begin{align}\label{phip4}
\phi' &= \frac{\z_1-\z_2}{4\k_1(d-1)W^{(3)}}=\frac{-2\z_3}{(d-1)(\z_1+\z_2)}\,,
\end{align}
where for simplicity in the notation and analysis it would be better to work with the following variables
\begin{subequations}
\begin{align}\label{zeta1}
\z_1 &= \sqrt{\z_2^2-8 \k_1 \z_3 W^{(3)}}\,,\\
 \z_2 &= -d\k_1{W'}^2+2((d-1)^2+d\k_1 W W'')\,,\\
\z_3 &= W' \big(\k_2 W^2-2 (d-1)^2 (d-1 -\k_1 V'')\big) + 6 (d-1)^2 \k_1 V'W''\,.
\label{zeta2}
\end{align}
\end{subequations}
Notice that the  expression \eqref{phip4} reduces to the equation \eqref{PhiEq} as $\k_1\rightarrow 0$. 
In the next step, we add two equations \eqref{GW1} and \eqref{GW2}. This gives a relation between superpotential and potential, which we need to solve for finding the solutions
\begin{align}\label{eom4d}
&\Big[4\k_1\z_2\z_3 W^{(3)}+2\k_1^2 \Big(16 (d-1)^4 V+d W\big(4 (d-1)^3 W-\k_2 W^3\nn \\
&-8 (d-1)^2 \k_1 V' W'\big)\Big) {W^{(3)}}^2-\z_2^3+\z_1\z_2^2\Big]/\big(\k_1 W^{(3)}\big)^2=0\,.
\end{align}
This is a polynomial equation for $W^{(3)}$ and to have a real solution for $W$ we should have the following condition
\be\label{cons1}
\z_2 \big(16 (d-1)^4 V+d W(4 (d-1)^3 W-\k_2 W^3-8 (d-1)^2 \k_1 V'W')\big)\geq 0\,.
\ee
This condition is precisely the reality condition for $\phi'(r)$ when we find it from equation \eqref{GW2}.
As we see, by keeping $\k_1=0$, we come back to the bound conditions that we found in section 3, \ie \eqref{nb}, \eqref{pb} and \eqref{bpm}. Therefore the above relation describes two possible bounds on the value of the superpotential $W$ and its derivatives. We should note that there is another condition for the reality of $\z_1$ in equation \eqref{zeta1} which we have used to find the condition \eqref{cons1}, this puts a constraint on the values of $W^{(3)}$. Consequently, this constraint together with \eqref{cons1} determine the allowed values of $W'$, $W''$ and $W^{(3)}$.

It is important to check another condition that connect with the curvature singularity. Similar to the $\k_1=0$ case, the curvature is given by
\be 
\mathcal{R}=
\frac{d}{d-1} \Big( \phi' W' -\frac{d+1}{4(d-1)} W^2  \Big)\,,
\ee
however, the singularity occurs when $\phi' W'$ diverges. If we suppose that $W'$ has a non-zero finite value then from equation of motion \eqref{GW1} we realize that the divergence of $\phi'$ happens at $\z_2=0$ or 
\be \label{SB}
-d\k_1{W'}^2+2((d-1)^2+d\k_1 W W'')=0\,.
\ee
Therefore, the $\z_2=0$ bound in \eqref{cons1} actually is a singularity bound.
This singularity can be avoided either when $W'\rightarrow 0$ faster than the divergence behavior of $\ph'$ or when $\ph'$ itself becomes finite. The latter happens if the \eqref{SB} bound intersect with the second bound in relation \eqref{cons1}, \ie
\be\label{betazz}
16 (d-1)^4 V+d W\big(4 (d-1)^3 W-\k_2 W^3-8 (d-1)^2 \k_1 V'W'\big)=0\,.
\ee
We can solve \eqref{SB} exactly. The curve of curvature singularities is given by
\be \label{SC}
W_S
=c(\ph-\ph_0)^2-\frac{(d-1)^2}{2c d \k_1}\,.
\ee
If we consider a solution of the superpotential around an arbitrary point $\ph$ then, we expect that this solution respects the above bounds. 
For example, let us suppose that RG flow starts or ends at the second bound \eqref{betazz}. If we insert \eqref{SC} into the \eqref{betazz} at this  point we can choose boundary conditions that fix the constants of integration, $\ph_0$ and $c$ in  \eqref{SC} and therefore fix the singularity curve. For a generic potential one choice would be $\ph_0=0$ and $c=-\frac{(d-1)\td{L}}{4d\k_1}$. Therefore, in this example the singularity curve is
\begin{subequations}
\begin{align} 
W_S&=\frac{2(d-1)}{\td{L}}-\frac{(d-1)\td{L}}{4d\k_1}\ph^2\,,\\
\td{L} &=\sqrt{\frac{-d(d-1)}{2V(0)}}\Big(1\pm\sqrt{1+\frac{4\k_2V(0)}{d(d-1)^2}}\Big)^\frac12\,.
\end{align}
\end{subequations}
The critical points of the RG flow can be found by analyzing the zero points of the beta function. We have the same definition of the beta function as in section 3
\be\label{beta4}
\beta=\frac{d\phi(r)}{dA(r)}
=\frac{\z_2-\z_1}{2\k_1 W W^{(3)}}
=\frac{\z_3}{4W(\z_1+\z_2)}\,.
\ee
So if we suppose a finite value for superpotential and its derivatives, the condition to have a critical point is coming from $\z_3=0$
\be\label{betaz}
W' \big(\k_2 W^2+2 (d-1)^2 (\k_1 V''-d+1)\big) + 6 (d-1)^2 \k_1 V'W''\big|_{\phi= \phi_c}=0\,.
\ee
If we insert this condition into the equation of motion \eqref{eom4d} and suppose that $\z_2\geq 0$, then find the equation \eqref{betazz} at the critical point $\phi=\phi_c$.
It means that this family of the critical points of \eqref{betaz} lives on the bound of \eqref{betazz} even when $\z_2=0$. We had observed this behavior already in section 3 when the line of $W=W_E$, which was the location of curvature singularities, met the upper or lower bounds. 
With the help of  \eqref{betazz} and \eqref{betaz} we can predict the conditions of existence of a critical point:

1) Local maxima and minima of the potential: Here, we assume that $W'=V'=0$, then $W^2=B_\pm^2$ gives the location of the bounds where $B_\pm$ are those in equation \eqref{bpm}.
This condition is similar to the previous section and corresponds to the critical points near the local maxima and minima of the potential. We should emphasize that the RG flow starts or ends on the bound \eqref{betazz} and always avoids hitting the singularity bound \eqref{SB}. 

2) Bounces: If we consider $V'\neq0$ it may have various situations:

$\bullet$ $W'=W''=0$ then the bound is located at $W_B^2=B_\pm^2$. 

$\bullet$ $W',W''\neq 0$ therefore  the bound is modified according to \eqref{betazz}. 

$\bullet$ $W'\neq 0$ but $W''=0$. There is a new value for superpotential at the bounce point
\be \label{bbb}
W_B=\pm(d-1)\sqrt{\frac{-2(\k_1 V''-d+1)}{\k_2}}\,.
\ee

All the above cases are not necessarily every possibility of the critical points because we supposed that every derivative of the superpotential is finite. For example, the last equality of \eqref{beta4} suggests that for a finite non-zero value of $\z_3$ when $W^{(3)}$ is diverging, the beta function tends to zero.

Finally, the direction of the RG flow is controlled by the sign of $W'(r)$
\be\label{WPN} 
\frac{dW(\phi(r))}{dr}=
W'\ph'=\frac{d-1}{d}\mathcal{R}+\frac{d+1}{4(d-1)}W^2
\ee
It is not clear that, in general, this function remains monotonic from one fixed point to another one along with the RG flow. Although we have supposed a negative value of the potential, it is not apparent that the curvature of the domain wall geometry remains negative or not as the holographic coordinate changes. The reason backs to the existence of the higher curvature correction terms in the Lagrangian. A conflict between these terms and potential may change the sign of the curvature. Nevertheless, we can put a bound on superpotential at each point if we want to have a monotonic function 
\begin{align} 
 |W|<\frac{2(d-1)}{L_c}\rightarrow \frac{dW}{dr}<0\,, \qquad
|W|>\frac{2(d-1)}{L_c}\rightarrow \frac{dW}{dr}>0\,,
\end{align}
where $L_c$ is a variable critical length at each point of the holographic coordinate $r$, and it defines by the curvature of space-time
\be \label{Lc}
L_c=2\sqrt{\frac{d(d+1)}{-\mathcal{R}}}\,.
\ee
So as far as the superpotential does not cross  the critical value $\frac{2(d-1)}{L_c}$ there is a c-function which may constructed from superpotential.
In the following sections, we will discuss the direction of the RG flow near the critical points. 
\subsection{Local maxima of potential}\label{41}
\begin{figure}[!ht]
\centering
\begin{subfigure}{0.48\textwidth}
\includegraphics[width=0.95\textwidth]{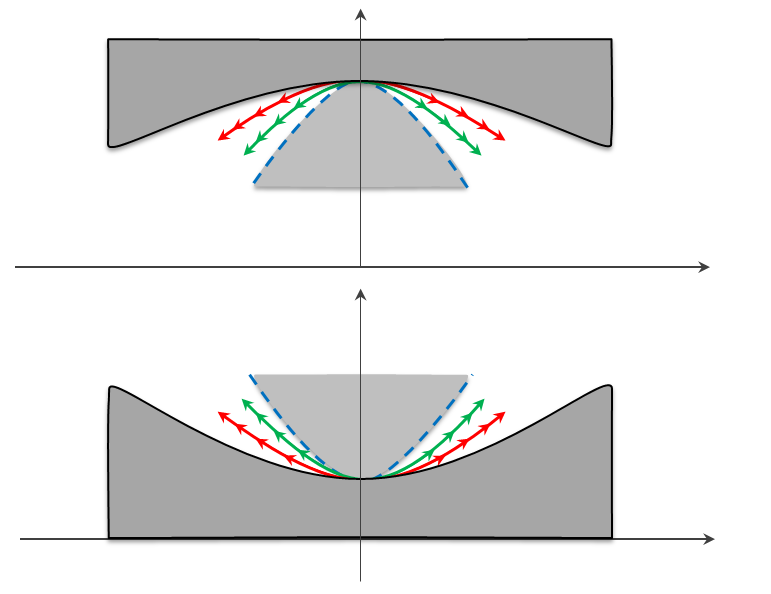}
\caption{\hspace*{6mm}}\label{f5}
\end{subfigure}
\centering
\begin{subfigure}{0.48\textwidth}
\includegraphics[width=0.95\textwidth]{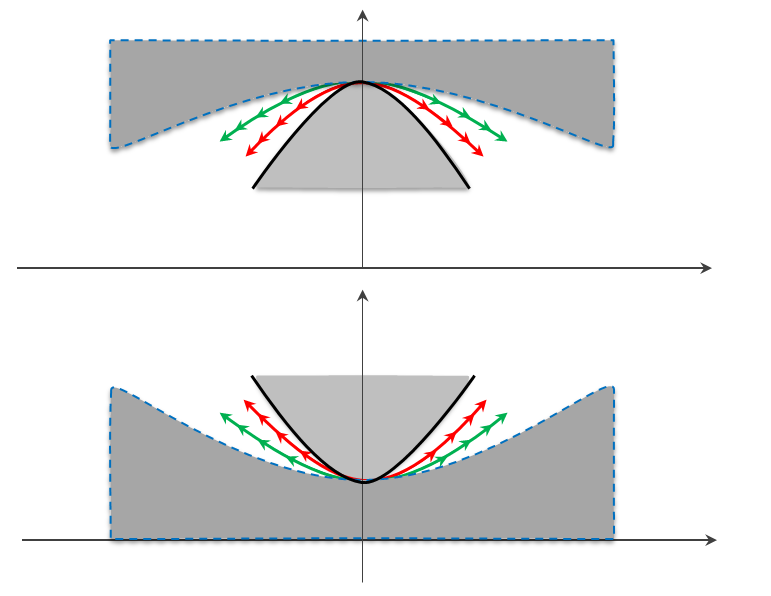}
\caption{\hspace*{5mm}}\label{f6}
\end{subfigure}
\centering
\begin{subfigure}{0.48\textwidth}\hspace*{1mm}
\includegraphics[width=0.95\textwidth]{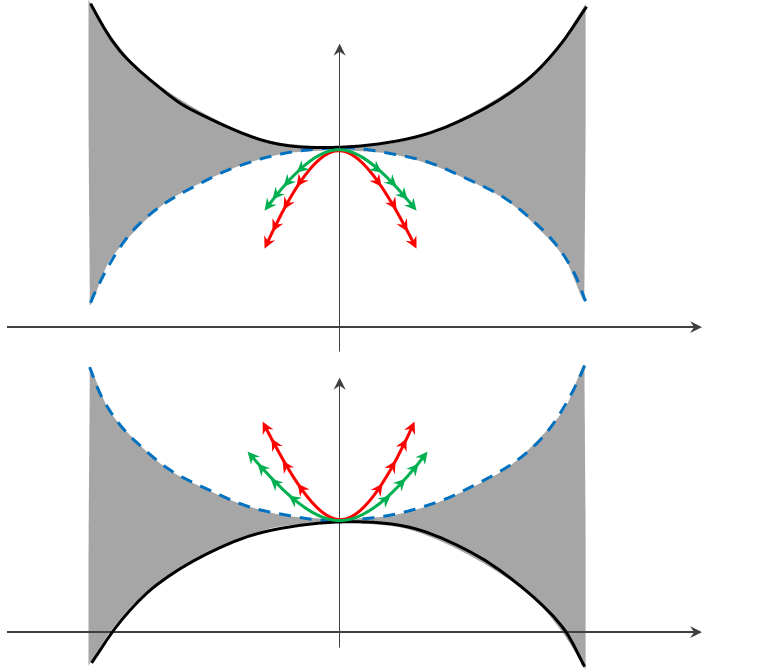}
\caption{\hspace*{5mm}}\label{f7}
\end{subfigure}
\centering
\begin{subfigure}{0.48\textwidth}
\includegraphics[width=0.95\textwidth]{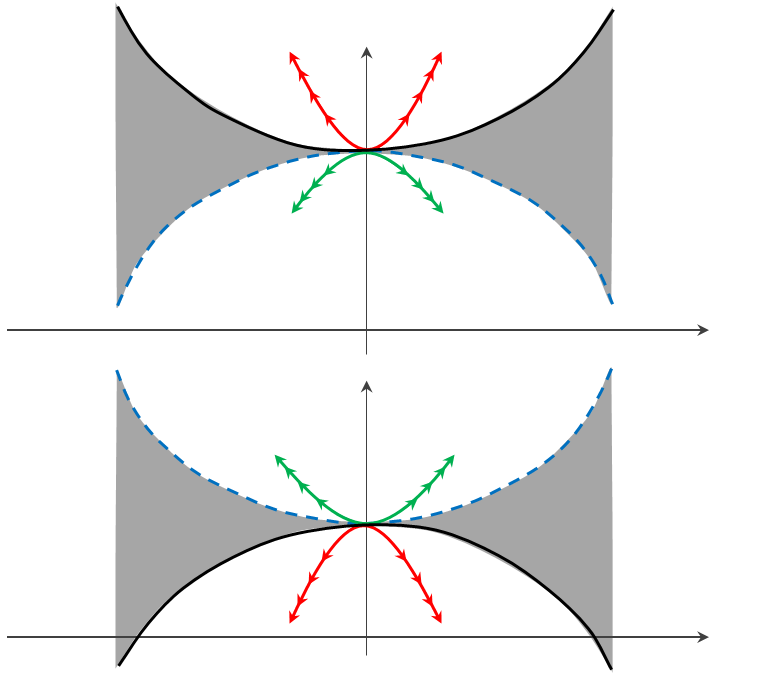}
\caption{\hspace*{5mm}}\label{f8}
\end{subfigure}
\centering
\begin{subfigure}{0.48\textwidth}\hspace*{-1.5mm}
\includegraphics[width=0.95\textwidth]{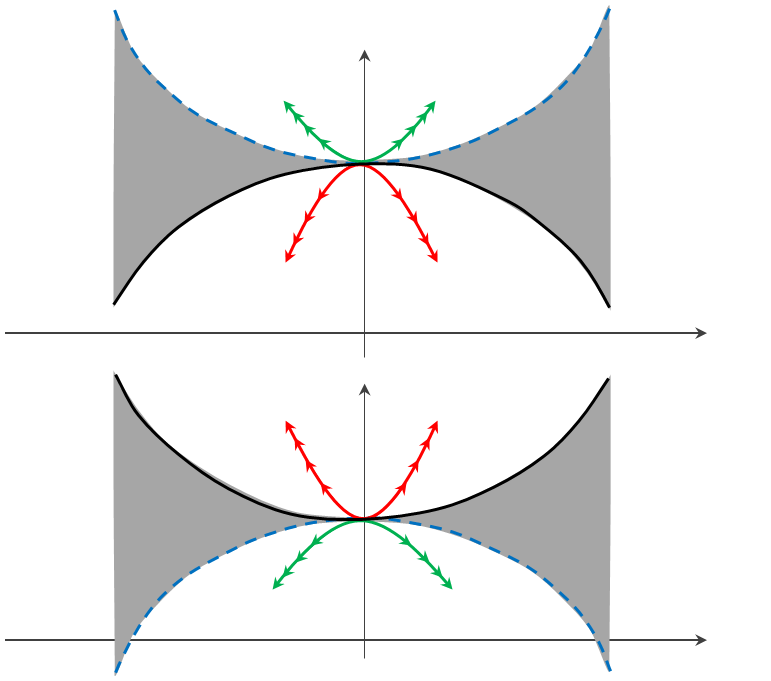}
\caption{\hspace*{5mm}}\label{f9}
\end{subfigure}
\centering
\begin{subfigure}{0.48\textwidth}\hspace*{-1.8mm}
\includegraphics[width=0.95\textwidth]{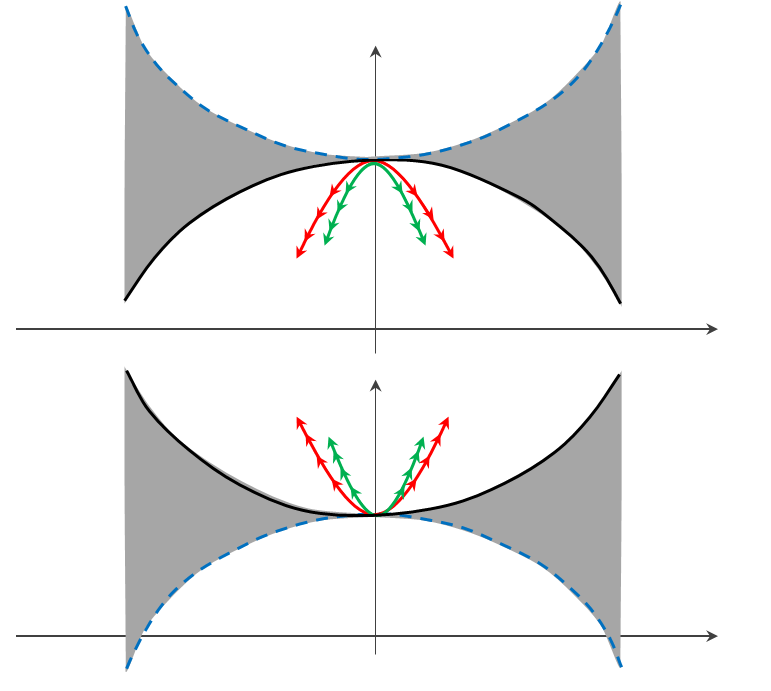}
\caption{\hspace*{5mm}}\label{f10}
\end{subfigure}
\caption{The generic behavior of superpotential near the local maxima of the potential. The green (red) curves belong to $W^+$ ($W^-$) branch. The shaded regions are the forbidden areas restricted by the blue bound (singularity curve) and the black (second) bound.}
\end{figure}

As a first type of the critical points, we examine the points near the local maxima of the potential. We consider again the following potential
\be 
V(\phi)=-\frac{d(d-1)}{L^2}-\frac12 m^2 \phi^2+O(\phi^3)\,,\qquad m^2>0\,.
\ee 
By solving the equation of motion \eqref{eom4d} we obtain
\begin{subequations}
\begin{align}
& \qquad\qquad\qquad W_{\pm}(\phi)=\frac{2(d-1)}{\tilde{L}}+\frac{\Delta_{\pm}}{2\tilde{L}}\phi^2+ \mathcal{C}_\pm \phi^{z_{\pm}}+\cdots\,,  \\ 
& \qquad\qquad\qquad \tilde{L}=\tilde{L}_\pm=\frac{L}{\sqrt{2}}\Big(1\pm \sqrt{1-\frac{4\k_2}{(d-1){L}^2}}\Big)^\frac12\,, \\ 
&\Delta_{\pm}
=\frac{2(d-1) \tilde{L}^4 m^2}{d\big((d-1)\tilde{L}^2-2\k_2\big)\pm \big(2\k_2-(4\k_1m^2+d-1)\tilde{L}^2\big)\sqrt{d^2-4m^2\tilde{L}^2}}\,.
\label{supma}
\end{align} 
\end{subequations}
These solutions reduce to \eqref{newparameters2}  when $\k_1=0$, and are real for $\k_2<\frac{(d-1)L^2}{4}$. The scalar field mass has an upper bound $m<\frac{d}{2\tilde{L}}$. Moreover for $\k_2<0$ just the $\td{L}=\tilde{L}_+$ is a valid choice \footnote{In the following computations we keep $\tilde{L}=\tilde{L}_\pm$ and analyze the results for both positive and negative values of $\k_2$ at the same time.}.
By solving the equation of motion to the leading term for $W_{+}$ solution we find three different values for $z_{+}$ as follows
\bea
u_0=\frac{2d}{d+\D_0}\,, \qquad u_{\pm}=\frac{d \pm \sqrt{d^2+\d }}{ d+\D_0}\,,
\eea
where we have defined the following parameters that control the behavior of the solutions 
\bea\label{dD0}
\d=4\frac{(d-1)\tilde{L}^2-2\k_2}{\k_1}\,,\qquad \D_0=\sqrt{d^2-4\tilde{L}^2 m^2}\,.
\eea
A simple algebraic analysis shows that we have the following regions for  $u_0$ and $u_\pm$. Note that only those regions give the leading terms that $z_+$ restricts to $2<z_+$ 
\begin{subequations}
\begin{align}
&u_{-}<0\,, \qquad\qquad 1<u_{0}<2\,,\qquad\quad\, u_0<u_{+}\,, \qquad\qquad\quad\, \d >0\,,\\
&0<u_{-}<1\,,\qquad  \frac{1}{2}<u_{+}\!<2\,, \qquad 1<u_{0}<2\,, \qquad -d^2<\d <0\,.
\end{align}
\end{subequations} 
Also the leading term for $W_{-}$ branch has the following values for $z_{-}$ 
\bea
v_0=\frac{2d}{d-\D_0}\,, \qquad v_{\pm}=\frac{d \pm \sqrt{d^2+\d }}{ d-\D_0}\,,
\eea
where these parameters are restricted
\begin{subequations}
\begin{align}
&v_{-}<0\,, \qquad\qquad\,\,\, 2<v_{0}<v_{+}\,, \qquad\quad\quad\quad\, \d >0\,,\\
&0<v_{-}<v_{+}<v_0\,, \qquad \,\, 2<v_{0}\,, \qquad -d^2<\d <0\,.
\end{align}
\end{subequations} 
According to the value of $\d$, we have different solutions for superpotential. We have listed the various regions for $\d$ in the first row of table \ref{t1}.
In this table we have defined
\begin{align} \label{deltapm}
\d_\pm=4\D_0(d\pm\D_0)\,.
\end{align}
The correct solutions of $W_\pm$ in each region are presented in the second and third rows of table \ref{t1}. 
\begin{table}[t]
\begin{center} 
\begin{tabular}{ |c||c|c|c|c|c| }
 \hline &&&&&\\[-1em]
 $\d$ interval & $\d\!<\!-\d_-\,, 
 \D_0\!<\!\frac{d}{2}$ & $\d\!<\!-\d_-\,, 
 \frac{d}{2}\!<\!\D_0$ & $-\d_-\!<\!\d\!<0$& $0<\!\d\!<\d_+$ & $\d_+\!<\d$ \\ &&&&&\\[-1em] \hline &&&&&\\[-1em]
 $W_+$ &  $W_+^2$& $W_+^2$& $W_+^2$& $W_+^2$ & $W_+^1$ \\  \hline &&&&&\\[-1em]
 $W_-$ & $W_-^1$ & $W_-^2$ &$W_-^3$ & $W_-^1$ & $W_-^1$ \\  \hline
\end{tabular}
\end{center}
\caption{Superpotential and corresponding regions near the local maxima of the potential.} \label{t1}
\end{table}
Every solution in this table is one of the following expressions
\begin{subequations}
\begin{align}\label{W1p}
& W^1_{+}=\frac{2(d-1)}{\tilde{L}}+\frac{\Delta_{+}}{2\tilde{L}}\phi^2+ \mathcal{C}_{+} \phi^{u_{+}}+\cdots\,,\\
&W^2_{+}=\frac{2(d-1)}{\tilde{L}}+\frac{\Delta_{+}}{2\tilde{L}}\phi^2+\cdots\,,\\
& W^1_{-}=\frac{2(d-1)}{\tilde{L}}+\frac{\Delta_{-}}{2\tilde{L}}\phi^2+ \mathcal{C}_{-} \phi^{v_{0}}+\cdots\,, \\
&W^2_{-}=\frac{2(d-1)}{\tilde{L}}+\frac{\Delta_{-}}{2\tilde{L}}\phi^2+ \mathcal{C}_{-} \phi^{v_{-}}+\cdots\,, \\
&W^3_{-}=\frac{2(d-1)}{\tilde{L}}+\frac{\Delta_{-}}{2\tilde{L}}\phi^2+ \mathcal{C}_{-} \phi^{v_{+}}+\cdots\,.\label{W3m}
\end{align}
\end{subequations} 
Here by using the relations \eqref{dD0} and \eqref{deltapm} we can rewrite $\D_\pm$ in equation \eqref{supma}  as
\be 
\D_\pm=\frac{-L^2\d (d\pm\D_0)}{2(L^2-2\tilde{L}^2)(\d\mp \d_{\pm})}\,. \\
\ee
We should also remember that for $\k_2>0$ both $\tilde{L}=\tilde{L}_+$ and $\tilde{L}=\tilde{L}_-$ are valid; therefore, we have two distinct sets of upper and lower bounds on the superpotential.
The superpotential solution must respect the bounds in \eqref{cons1}. Since we have assumed    $V'=0$, the shape of the second bound \eqref{betazz} is given by $W=B_\pm$  in \eqref{bpm}. 
To analyze and draw the shape of the superpotential easier, it would be better to find the shape of the bounds approximately. If we insert a series solution similar to the superpotential, into the equation \eqref{cons1} we will get the following results
\begin{subequations}
\begin{align} \label{bmax}
&W_b=\frac{2(d-1)}{\tilde{L}}+\frac{\D_b}{2\tilde{L}}\phi^2+\cdots\,,\qquad \D_b=\D_{(1,2)}\,,\\ 
&\D_1=-\frac{(d-1)\tilde{L}^2}{2d \k_1}\,,\qquad\D_2=\frac{(d-1)\tilde{L}^4 m^2}{d(-2\k_2+\tilde{L}^2(d-1+2\k_1 m^2))}\,.
\end{align}
\end{subequations}
To simplify our analysis we write
\begin{align}
\D_1=\frac{L^2\d}{8d(L^2-2\tilde{L}^2)}\,,\quad
\D_2=\frac{L^2\d \d^*}{8d(L^2-2\tilde{L}^2)(\d- \d^*)}\,, \qquad \d^*=2(\D_0^2-d^2)\,.
\end{align}
Since $0<\D_0<d$, we have $\d^*<-\d_-<0<\d_+$. Moreover,  $0 < \td{L}_- \leq \frac{L}{\sqrt{2}} \leq \td{L}_+ < L$. Therefore in various regions of $\d$ (the first column of table \ref{t2}), the signs of $\D_b$'s  determine the region of forbidden area, bounded by the \eqref{bmax} curves. 
\begin{table}[t]
\begin{center} 
\begin{tabular}{|c|c|c|c|}
 \hline &&&\\[-1em]&&&\\[-1em]
 $\d$ interval & $\tilde{L}=\tilde{L}_-$ & $\tilde{L}=\tilde{L}_+$ & Fig. \\ &&&\\[-1em] \hline &&&\\[-1em]
 $(-d^2,2\d^*)$ & $0>\D_2>\D_->\D_+>\D_1$ & $\D_1>\D_+>\D_->\D_2>0$ &\ref{f5}
 \\ &&&\\[-1em] \hline &&&\\[-1em]
 $(2\d^*,\d^*)$ & $0>\D_1>\D_+>\D_->\D_2$ & $\D_2>\D_->\D_+>\D_1>0$ &\ref{f6}
 \\ &&&\\[-1em] \hline &&&\\[-1em]
 $(\d^*,-\d_-)$& $\D_2>0>\D_1>\D_+>\D_-$ & $\D_->\D_+>\D_1>0>\D_2$ &\ref{f7}
 \\ &&&\\[-1em] \hline &&&\\[-1em]
 $(-\d_-,0)$ & $\D_->\D_2>0>\D_1>\D_+$ & $\D_+>\D_1>0>\D_2>\D_-$ &\ref{f8}
 \\ &&&\\[-1em] \hline &&&\\[-1em]
 $(0,\d_+)$& $\D_+>\D_1>0>\D_2>\D_-$ & $\D_->\D_2>0>\D_1>\D_+$ &\ref{f9}
 \\ &&&\\[-1em] \hline &&&\\[-1em]
 $(\d_+,\infty)$ & $\D_1>0>\D_2>\D_->\D_+$ & $\D_+>\D_->\D_2>0>\D_1$ & \ref{f10} \\  \hline 
\end{tabular}
\end{center}
\caption{The shape of upper and lower bounds depends on the signs of $\D_b$. The orientation of $W_{\pm}$ curves also can be read from this table.} 
\label{t2}
\end{table}
The shape of $W_\pm$ curves also depends on the relation between $\D_\pm$ with $\D_b$. 
We have summarized these information in the  second and third columns of table \ref{t2} for $\tilde{L}=\tilde{L}_\pm$. Note that for the upper bound $\tilde{L}=\tilde{L}_-$ and for lower bound $\tilde{L}=\tilde{L}_+$. 
For every region, we can draw a generic behavior of the superpotential. We have addressed every related configuration in the last column of the table \ref{t2} to figures \ref{f5} to \ref{f10}. To find which $W_\pm$ corresponds to which figure, we should use the table \ref{t1} according to the region of $\d$.
To compute the scalar field, we insert each function of $W_\pm$ into the equation \eqref{phip4} and then solve the differential equation
\begin{subequations}
\begin{align}
\phi^1_{+}(r)&=\phi_+ e^{\frac{d}{\tilde{L}u_0} r}+\frac{\alpha(u_0,u_+,u_-) \mathcal{C}_{+}\tilde{L} u_0 }{d u_{+}} (\phi_{+}e^{\frac{d}{\tilde{L}u_0} r})^{u_{+}+1}+\cdots\,, \\
\phi^2_{+}(r)&=\phi_+ e^{\frac{d}{\tilde{L}u_0} r}+\cdots\,, \\
\phi^1_{-}(r)&=\phi_- e^{\frac{d}{\tilde{L} v_0} r}+\frac{\alpha\, \mathcal{C}_{-}\tilde{L} v_0 }{d(v_0-2)} (\phi_- e^{\frac{d}{\tilde{L} v_0} r})^{v_0-1}+\cdots\,, \\[-2pt]
\phi^2_{-}(r)&=\phi_- e^{\frac{d}{\tilde{L}v_0} r}+\frac{\a(v_0,v_-,v_+) \mathcal{C}_{-}\tilde{L} v_0 }{d v_{-}} (\phi_- e^{\frac{d}{\tilde{L} v_0} r})^{v_- +1}+\cdots\,,
\\[-2pt]
\phi^3_{-}(r)&=\phi_- e^{\frac{d}{\tilde{L}v_0} r}+\frac{\a(v_0,v_+,v_-) \mathcal{C}_{-}\tilde{L} v_0 }{d v_{+}} (\phi_- e^{\frac{d}{\tilde{L} v_0} r})^{v_+ +1}+\cdots\,,
\end{align}
\end{subequations}
\pagebreak
\pagebreak

where in above relations we have used the following definitions
\begin{subequations}
\begin{align}\label{alphas}
 \alpha(a,b,c) &=\frac{1}{2d(d-1) \k_1 (4 - 4 a + b c)^2}\Big(d^2 \k_1 (4 - 4 a+ b c) (2 - 3 b + b^2)\nn \\
&+  2  \k_2 a \big(8 + 2 (4 + c) b + (c -6) b^2 + 2 b^3 + 2 a (b^2- 3b -4 )\big)\Big)\,,\\
\a &=\frac{d^2 \k_1 v_0 v_+ v_- (4 - 2 v_0 + v_+ v_- )}{(4 v_0 -v_+ v_- -4  ) (2 \k_2 v_0^2 - d^2 \k_1 v_+ v_-)}\,.
\end{align}
\end{subequations}

In the last step, we can find the scale factors associated with the above solutions
\begin{subequations}
\begin{align}
A^1_+(r) &=-\frac{r-r_*}{\tilde{L}}+\frac{u_0 \Delta_+ }{8d(1-d)}\big(\phi_+ e^{\frac{d}{u_0\tilde{L}}r}\big)^2+\frac{\mathcal{C}_+ \tilde{L} u_0}{2d(1-d)u_+}\big(\phi_+ e^{\frac{d}{u_0\tilde{L}}r}\big)^{u_+}+\cdots\,, \\
A^2_+(r) &=-\frac{r-r_*}{\tilde{L}}+\frac{u_0 \Delta_+ }{8d(1-d)}\big(\phi_+ e^{\frac{d}{u_0\tilde{L}}r}\big)^2+\cdots\,, \\[-4pt] 
A^1_-(r) &=-\frac{r-r_*}{\tilde{L}}+\frac{v_0 \Delta_- }{8d(1-d)}\big(\phi_- e^{\frac{d}{v_0\tilde{L}}r}\big)^2+\frac{\mathcal{C}_- \tilde{L} (1+\frac{v_0 \a \Delta_-}{d (v_0-2)})}{2d(1-d)}\big(\phi_- e^{\frac{d}{v_0\tilde{L}}r}\big)^{v_0}+\cdots\,, \\
A^2_-(r) &=-\frac{r-r_*}{\tilde{L}}+\frac{v_0 \Delta_- }{8d(1-d)}\big(\phi_- e^{\frac{d}{v_0\tilde{L}}r}\big)^2+\frac{\mathcal{C}_- \tilde{L} v_0}{2d(1-d)v_-}\big(\phi_- e^{\frac{d}{v_0\tilde{L}}r}\big)^{v_-}+\cdots\,,\\
A^3_-(r) &=-\frac{r-r_*}{\tilde{L}}+\frac{v_0 \Delta_- }{8d(1-d)}\big(\phi_- e^{\frac{d}{v_0\tilde{L}}r}\big)^2+\frac{\mathcal{C}_- \tilde{L} v_0}{2d(1-d)v_+}\big(\phi_- e^{\frac{d}{v_0\tilde{L}}r}\big)^{v_+}+\cdots\,.
\end{align}
\end{subequations}
As a result, the local maxima is a UV fixed point at $r\rightarrow -\infty$. This is clear also from the computation of the beta function near the extremum point
\begin{align}
\b^{1,2}_{+}(r)=-\frac{d}{u_0}\phi_+e^{\frac{d}{u_0\tilde{L}}r}+\cdots\,,\qquad
\beta^{1,2,3}_{-}(r)=-\frac{d}{v_0}\phi_{-}e^{\frac{d}{v_0\tilde{L}}r}+\cdots\,,
\end{align}
where all the beta functions are vanishing at the fixed point. This is exactly what we expect from equation \eqref{betaz} where for all solutions of the superpotential at the critical point, $W'$ vanishes and $W''$ is finite. 
\subsection{Local minima of potential}\label{42}
For the local maxima of the potential, we consider again the following potential
\be 
V(\phi)=-\frac{d(d-1)}{L^2}+\frac12 m^2 \phi^2+O(\phi^3)\,,\qquad m^2>0\,.
\ee 
The solutions of the equation of motion  \eqref{eom4d} are similar to the local maxima solutions with a change in the sign of $m^2$, specifically
\bea
\Delta_{\pm}
=\frac{-2(d-1) \tilde{L}^4 m^2}{d\big((d-1)\tilde{L}^2-2\k_2\big)\pm \big(2\k_2+(4\k_1m^2-d+1)\tilde{L}^2\big)\sqrt{d^2+4m^2\tilde{L}^2}}\,.
\eea  
The computation of the leading term for $W_{+}$ branch leads to three different values for $z_{+}$ with different ranges
\begin{subequations}\label{jj1}
\begin{align}
&u_0=\frac{2d}{d+\D_0}\,, \qquad\qquad\,\,\,\, u_{\pm}=\frac{d \pm \sqrt{d^2+\d }}{ d+\D_0}\,,\\
&u_{-}<0<u_{0}<1\,,\qquad\quad u_0<u_{+}\,, \qquad\qquad\qquad \,\,\,\d>0\,,\\
&0<u_{-}<u_{+}<u_0<1\,, \qquad\qquad\qquad\qquad -d^2<\d<0\,,
\end{align}
\end{subequations}
and we have defined
\bea
\d=4\frac{(d-1)\tilde{L}^2-2\k_2}{\k_1}\,,\qquad
\D_0=\sqrt{d^2+4\tilde{L}^2 m^2}\,.
\eea
The leading terms of $W_{-}$ branch have also three values for $z_{-}$ with the following constraints 
\begin{subequations}
\begin{align}
&v_0=\frac{2d}{d-\D_0}\,, \qquad v_{\pm}=\frac{d \pm \sqrt{d^2+\d }}{ d-\D_0}\,,\\
&v_{+}<v_0<0<v_{-}\,, \qquad\qquad\quad\,\, \d >0\,,\\
&v_{0}<v_- <v_+ <0\,,\,\qquad\, -d^2<\d <0\,.
\end{align}
\end{subequations} 
There exist three regions for $\d$ with a specific solution of $W_\pm$ in each region. These are shown in the first row of table \ref{t3} where $\d_\pm$ have the same definition as \eqref{deltapm}.
\begin{table}[t]
\begin{center} 
\begin{tabular}{ |c||c|c|c| }
 \hline &&&\\[-1em]
 Interval &  $-d^2\!<\!\d\!<\!-\d_-$ & $-\d_-\!<\!\d\!<\d_+$ & $\d_+\!<\d$ \\ &&&\\[-1em] \hline &&&\\[-1em]
 $W_+$ & $W_+^2$&  $W_+^2$ & $W_+^1$ \\  \hline &&&\\[-1em]
 $W_-$ & $W_-^2$ & $W_-^1$ & $W_-^1$ \\  \hline
\end{tabular}
\end{center}
\caption{Superpotential and corresponding regions near the local minima of the potential.} \label{t3}
\end{table}

The $W_\pm$ solutions are presented in the second and third rows of table \ref{t3}.
Every solution in this table has one of the following values
\begin{subequations}
\begin{align}\label{W1pn}
& W^1_{+}=\frac{2(d-1)}{\tilde{L}}+\frac{\Delta_{+}}{2\tilde{L}}\phi^2+ C_{+} \phi^{u_{+}}+\cdots\,,\\
&W^2_{+}=\frac{2(d-1)}{\tilde{L}}+\frac{\Delta_{+}}{2\tilde{L}}\phi^2+\cdots\,,\\
& W^1_{-}=\frac{2(d-1)}{\tilde{L}}+\frac{\Delta_{-}}{2\tilde{L}}\phi^2+ C_{-} \phi^{v_{-}}+\cdots\,,\\
&W^2_{-}=\frac{2(d-1)}{\tilde{L}}+\frac{\Delta_{-}}{2\tilde{L}}\phi^2+\cdots\,.\label{W2mn}
\end{align}
\end{subequations} 
To draw the RG flow curves near the local minima of the potential, similar to the local maxima, we find the shape of the bounds from equation \eqref{cons1}. Here again, at the critical point, the upper or lower bounds are given by $B_\pm$ in equation \eqref{bpm}. A little away from this point, we should solve the constraint \eqref{cons1}.
If we insert a  series solution into the equation \eqref{cons1}, we obtain the following results
\begin{subequations}
\begin{align} \label{bmin}
&W_b=\frac{2(d-1)}{\tilde{L}}+\frac{\D_b}{2\tilde{L}}\phi^2+\cdots\,,\qquad \D_b=\D_{(1,2)}\\ 
&\D_1=-\frac{(d-1)\tilde{L}^2}{2d \k_1}\,,\qquad\D_2=\frac{-(d-1)\tilde{L}^4 m^2}{d(-2\k_2+\tilde{L}^2(d-1-2\k_1 m^2))}\,.
\end{align}
\end{subequations}
To simplify our analysis let's write
\begin{subequations}
\begin{align}
&\D_1=\frac{L^2\d}{8d(L^2-2\tilde{L}^2)}\,,\qquad\qquad\quad
\D_2=\frac{-L^2\d \d^*}{8d(L^2-2\tilde{L}^2)(\d+ \d^*)}\,,\\
&\D_\pm=\frac{-L^2\d (d\pm\D_0)}{2(L^2-2\tilde{L}^2)(\d\mp \d_{\pm})}\,,\qquad \d_\pm=4\D_0(d\pm\D_0)\,, \qquad
\d^*=2(d^2-\D_0^2)\,,
\end{align}
\end{subequations}
where $0\!<\!-\d^*\!<\!-\d_-\!<\!-2\d^*\!<\!\d_+$.
For various values of $\d$ (the first column of table \ref{t4}), the signs of $\D_b$'s  determine the region of forbidden areas bounded by the \eqref{bmin} curves. The orientation of $W_\pm$ curves also depend on the relation between $\D_\pm$ with $\D_b$. We have summarized these information in the  second and third columns of table \ref{t4} for $\tilde{L}=\tilde{L}_\pm$. 

For every region, we can draw a generic behavior of superpotential. We have addressed the related figures in the last column of table \ref{t4}. To find which $W_\pm$ corresponds to each figure we should use table \ref{t3} according to the region of $\d$. 
\begin{table}[t]
\begin{center} 
\begin{tabular}{ |c|c|c|c|}
 \hline &&&\\[-1em]&&&\\[-1em]
 $\d$ interval & $\tilde{L}=\tilde{L}_-$ & $\tilde{L}=\tilde{L}_+$ & Fig. \\ &&&\\[-1em] \hline &&&\\[-1em]
 $(-d^2,0)$ & $\D_2>\D_->0>\D_+>\D_1$ & $\D_1>\D_+>0>\D_->\D_2$ &\ref{f11}
 \\ &&&\\[-1em] \hline &&&\\[-1em]
 $(0,-\d^*)$ & $\D_1>\D_+>0>\D_->\D_2$ & $\D_2>\D_->0>\D_+>\D_1$ &\ref{f12}
 \\ &&&\\[-1em] \hline &&&\\[-1em]
 $(-\d^*,-\d_-)$& $\D_2>\D_1>\D_+>0>\D_-$ & $\D_->0>\D_+>\D_1>\D_2$ &\ref{f13}
 \\ &&&\\[-1em] \hline &&&\\[-1em]
 $(-\d_-,-2\d^*)$ & $\D_->\D_2>\D_1>\D_+>0$ & $0>\D_+>\D_1>\D_2>\D_-$ &\ref{f14}
 \\ &&&\\[-1em] \hline &&&\\[-1em]
 $(-2\d^*,\d_+)$& $\D_+>\D_1>\D_2>\D_->0$ & $0>\D_->\D_2>\D_1>\D_+$ &\ref{f15}
 \\ &&&\\[-1em] \hline &&&\\[-1em]
 $(\d_+,\infty)$ & $\D_1>\D_2>\D_->0>\D_+$ & $\D_+>0>\D_->\D_2>\D_1$ &\ref{f16}
 \\  \hline
\end{tabular}
\end{center}
\caption{The shape of upper and lower bounds depends on the signs of $\D_b$. The orientation of $W_{\pm}$ curves also can be read from this table.} \label{t4}
\end{table}

By insertion of the values of $W_\pm$ from \eqref{W1pn} to \eqref{W2mn} into the equation \eqref{phip4} and solving the differential equation we obtain the scalar field solution
\begin{subequations}
\begin{align}
\phi^1_{+}(r)&=\phi_+ e^{\frac{d}{\tilde{L}u_0} r}+\frac{\alpha(u_0,u_+,u_-) \mathcal{C}_{+}\tilde{L} u_0 }{d u_{+}} (\phi_{+}e^{\frac{d}{\tilde{L}u_0} r})^{u_{+}+1}+\cdots\,,\\
\phi^2_{+}(r)&=\phi_+ e^{\frac{d}{\tilde{L}u_0} r}+\cdots\,,\\
\phi^1_{-}(r)&=\phi_- e^{\frac{d}{\tilde{L} v_0} r}+\frac{\a(v_0,v_-,v_+) \mathcal{C}_{-}\tilde{L} v_0 }{dv_-} (\phi_- e^{\frac{d}{\tilde{L} v_0} r})^{v_-+1}+\cdots\,, \\
\phi^2_{-}(r)&=\phi_- e^{\frac{d}{\tilde{L}v_0} r}+\cdots\,,
\end{align}
\end{subequations} 
where the coefficients can be read from equation \eqref{alphas}. In addition, any scale factor associated to each superpotential is given by
\begin{subequations}
\begin{align}
A^1_+(r) &=-\frac{r-r_*}{\tilde{L}}+\frac{u_0 \Delta_+ }{8d(1-d)}\big(\phi_+ e^{\frac{d}{u_0\tilde{L}}r}\big)^2+\frac{\mathcal{C}_+ \tilde{L} u_0}{2d(1-d)u_+}\big(\phi_+ e^{\frac{d}{u_0\tilde{L}}r}\big)^{u_+}+\cdots\,, \\
A^2_+(r) &=-\frac{r-r_*}{\tilde{L}}+\frac{u_0 \Delta_+ }{8d(1-d)}\big(\phi_+ e^{\frac{d}{u_0\tilde{L}}r}\big)^2+\cdots\,, \\
A^1_-(r) &=-\frac{r-r_*}{\tilde{L}}+\frac{v_0 \Delta_- }{8d(1-d)}\big(\phi_- e^{\frac{d}{v_0\tilde{L}}r}\big)^2+\frac{\mathcal{C}_- \tilde{L} v_0}{2d(1-d)v_-}\big(\phi_- e^{\frac{d}{v_0\tilde{L}}r}\big)^{v_-}+\cdots\,,\\
A^2_-(r) &=-\frac{r-r_*}{\tilde{L}}+\frac{v_0 \Delta_- }{8d(1-d)}\big(\phi_- e^{\frac{d}{v_0\tilde{L}}r}\big)^2+\cdots\,.
\end{align}
\end{subequations}
Since $u_0>0$ and $v_0<0$, at the local minima,  the $W_+$ branch tends to a UV fixed point as $r\rightarrow -\infty$, in contrary, this point is an IR fixed point for $W_-$ branch as $r\rightarrow +\infty$. 

The behavior of the beta function near the UV/IR fixed point is as follow
\begin{align}
\b^{1,2}_{+}(r)=-\frac{d}{u_0}\phi_+e^{\frac{d}{u_0\tilde{L}}r}+\cdots\,,\qquad
\beta^{1,2}_{-}(r)=-\frac{d}{v_0}\phi_{-}e^{\frac{d}{v_0\tilde{L}}r}+\cdots\,,
\end{align}
where the both functions vanish at the fixed point. This is an expected result from equation \eqref{betaz}, where at the critical point, $W'$ is zero while $W''$ is finite. 
\begin{figure}[!ht]
\centering
\begin{subfigure}{0.48\textwidth}
\includegraphics[width=0.95\textwidth]{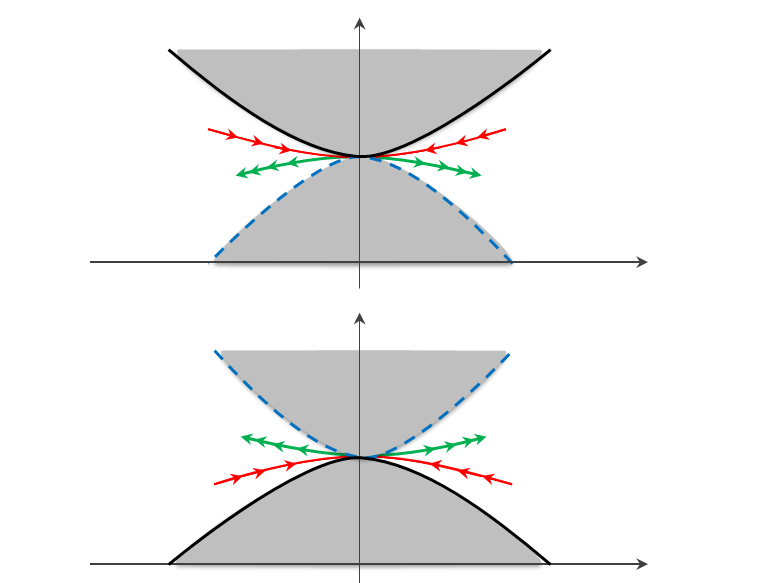}
\caption{\hspace*{7mm}}\label{f11}
\end{subfigure}
\centering
\begin{subfigure}{0.48\textwidth}
\includegraphics[width=0.95\textwidth]{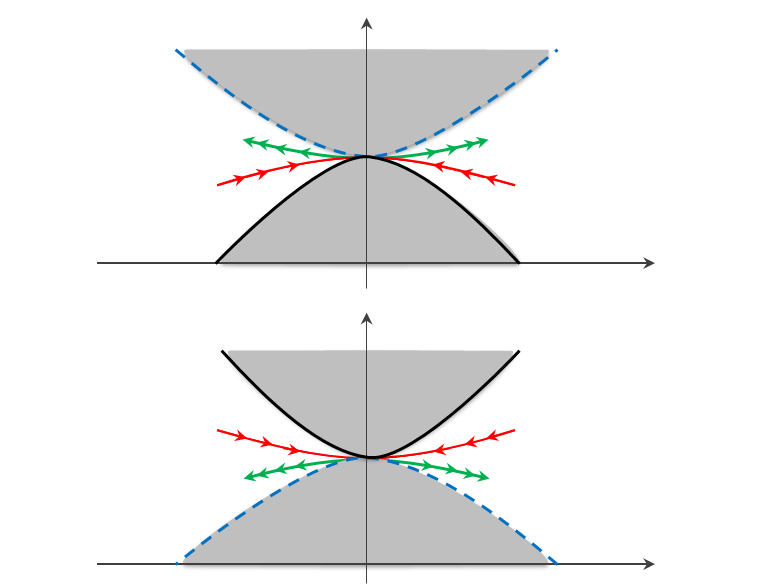}
\caption{\hspace*{7mm}}\label{f12}
\end{subfigure}
\centering
\begin{subfigure}{0.48\textwidth}\hspace*{-2.4mm}
\includegraphics[width=0.95\textwidth]{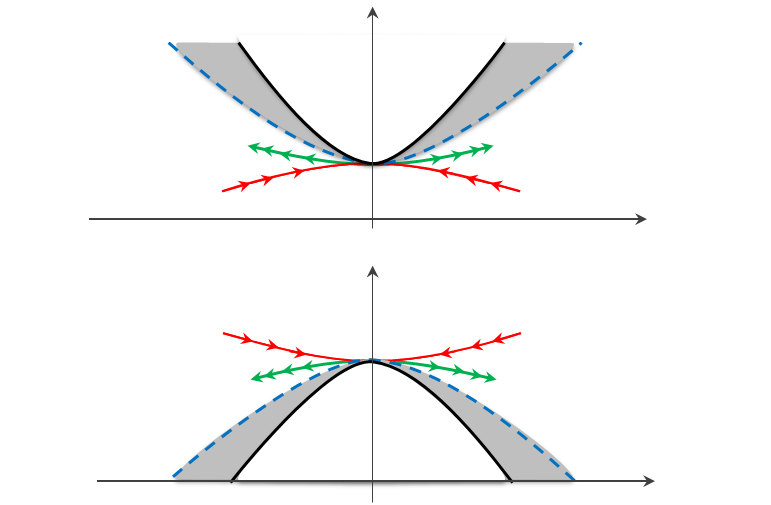}
\caption{\hspace*{7mm}}\label{f13}
\end{subfigure}
\centering
\begin{subfigure}{0.48\textwidth}\hspace*{-1.5mm}
\includegraphics[width=0.95\textwidth]{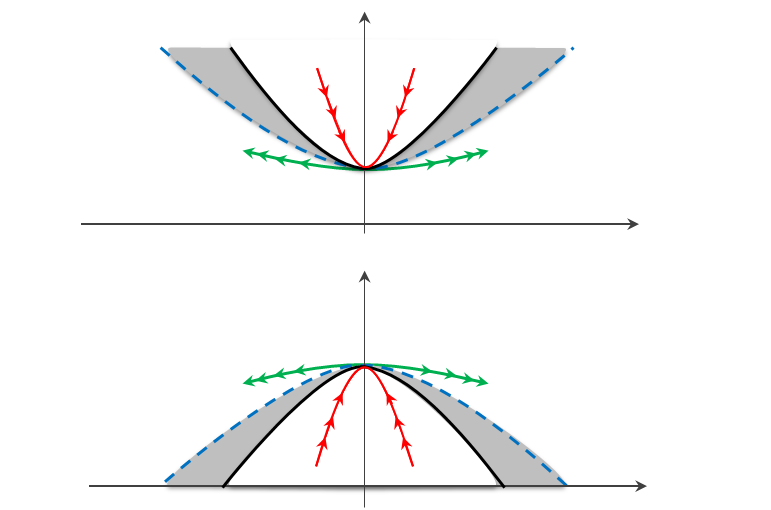}
\caption{\hspace*{7mm}}\label{f14}
\end{subfigure}
\centering
\begin{subfigure}{0.48\textwidth}\hspace*{-2.5mm}
\includegraphics[width=0.95\textwidth]{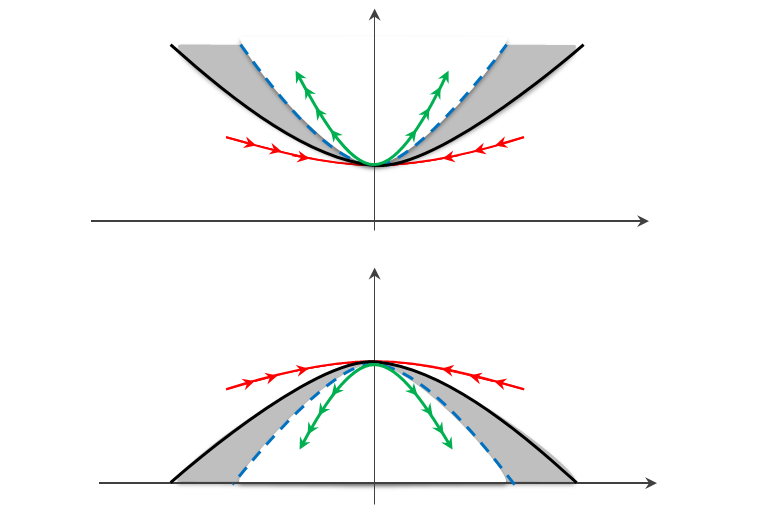}
\caption{\hspace*{7mm}}\label{f15}
\end{subfigure}
\centering
\begin{subfigure}{0.48\textwidth}\hspace*{-2.7mm}
\includegraphics[width=0.95\textwidth]{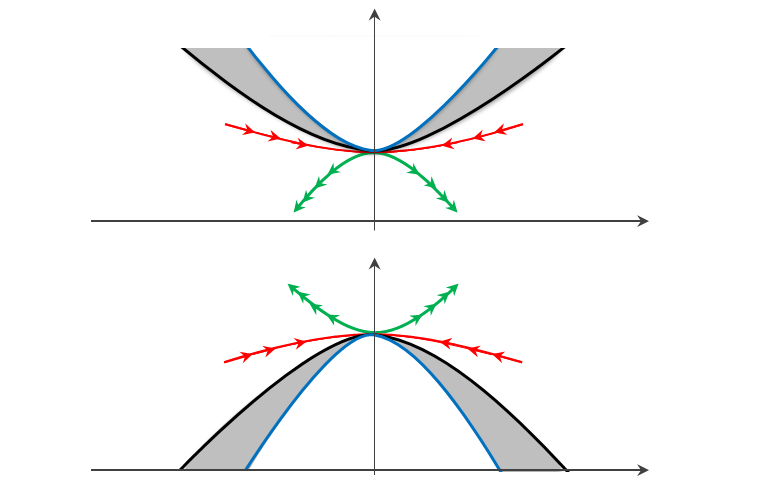}
\caption{\hspace*{7mm}}\label{f16}
\end{subfigure}
\caption{The generic behavior of superpotential near the local minima of the potential. The green (red) curves belong to $W^+$ ($W^-$) branch.}
\end{figure}
\subsection{Bounces}\label{43}
To have a general analysis of other possible solutions, let's consider the following ansatz near a critical point $\phi=\phi_B$ as a starting point
\begin{subequations}
\begin{align}
V(\ph) &= \sum_{n=0} V_n(\ph-\ph_B)^n\,,\\
W(\ph) &= W_B+ \sum_{i=0}^{5}\mathcal{C}_i(\phi-\ph_B)^\frac{i+1}{2}+\mathcal{C}_z(\ph-\ph_B)^z+\cdots\,,\label{Bo4}
\end{align} 
\end{subequations}
where we have assumed that $V'\neq 0$ at $\phi_B$  which means that the potential is no longer located near a local minimum or maximum of the potential. Moreover, although in this ansatz $W'$ diverges at $\phi_B$ but one can check that at the same time $\phi'\rightarrow 0$, so the curvature remains finite.
If we put the above ansatz into the equation of motion \eqref{eom4d} then we can read the unknown coefficients. For example the first three coefficients are
\begin{subequations}
\begin{align}
 \mathcal{C}_1 &=\mathcal{C}_0 \Big( \frac{\mathcal{C}_0}{4 W_B} + \frac{ d W_B}{3 \sqrt{2V_1} (d -1)} \Big) + \frac{16 (d-1)^4 V_0 + 4 d(-1 + d)^3 W_B^2 -  d \kappa_2 W_B^4}{8 d(d-1)^2 \kappa_1 V_1 W_B}\,,\\
\mathcal{C}_2 &= \mathcal{C}_0 \Big( \frac{12 (d-1)^2 (2d-2 - \k_1 V_2) + (d^2 \k_1 - 12 \k_2) W_B^2}{72 (d-1)^2 \k_1 V_1} + \frac{\mathcal{C}_0 d}{4 \sqrt{2V_1} (d-1)} \Big)\,,\\
 \mathcal{C}_3 &=- \frac{\mathcal{C}_0^2 (d^2 \k_1 + 20 \k_2) W_B - 20 \mathcal{C}_1 \bigl( 2 (d-1)^2 ( d-1 - 2 \k_1 V_2) - \k_2 W_B^2\bigr)}{240 (d-1)^2 \k_1 V_1} \\
    &+ \frac{\mathcal{C}_0 d\Big(108 \mathcal{C}_1 (d-1)^2 \k_1 V_1 - 6 (d-1)^2 ( 5(d-1) - 4 \kappa_1 V_2) W_B- ( d^2 \kappa_1 - 15\k_2) W_B^3 \Big)}{540 \sqrt{2 V_1^3} (d-1)^3 \k_1} \,, \nn
\end{align}
\end{subequations}
where we have supposed that $\mathcal{C}_0\neq0$ in finding the above results.
Here for $\ph>\ph_B$ we assumed that $V_1>0$, otherwise for $\ph<\ph_B$, we should consider $V_1<0$. It is related to how the bounce is reaching the bound from the left or the right.

In this general solution, although all derivatives of $W$ diverge at the critical point, the bounce solution respects the condition \eqref{cons1}. Moreover, we can write all the coefficients in terms of $\mathcal{C}_1$ and $\mathcal{C}_3$ or $\ph$ and $\ph^2$ coefficients, in this way, the location of the bound, $W_B$, is specified by two boundary conditions.

Now we can compute the scalar field and scale factor and we get
\begin{subequations}
\begin{align}
\phi(r) &=\phi_B+\frac12 V_1 (r-r_B)^2+\mathcal{O}(r-r_B)^3\,,\\
A(r) &=A_B-\frac{W_B}{2(d-1)}(r-r_B)-\frac{\mathcal{C}_0\sqrt{2V_1}}{8(d-1)}(r-r_B)^2+\cdots\,.
\end{align}
\end{subequations}
This shows that their values are independent of two branches of the superpotential.
The behavior of the beta function near the bounce point is given by
\be 
\beta(\ph)=-sign(\mathcal{C}_0)\frac{2(d-1)\sqrt{2V_1}}{W_B} (\ph-\ph_B)^\frac12+\mathcal{O}(\ph-\ph_B)\,.
\ee
We can trace the direction of RG flow by computing
\be 
\frac{d W(r)}{dr}=\sqrt{\frac{V_1}{2}} \mathcal{C}_0 +\cdots\,.
\ee
For a fixed value of $\mathcal{C}_0$, as the holographic coordinate changes from values below the $r=r_B$ to the values above it, the direction of the RG flow does not change. It means that $r_B$ is a bounce point.

As we told, in the above solutions, we have considered $\mathcal{C}_0\neq 0$. To find other possible bounces with other boundary conditions, we should solve the equations of motion from the beginning. These solutions are those that we indicated for the bounce points above the equation \eqref{WPN}. In the following sections, we find these solutions.
\subsubsection{\texorpdfstring{$W'= W''=0$}{W'=W''=0}}
As we mentioned earlier, the boundary location is determined when we impose boundary conditions on $W'$ and $W''$. To maintain on the bound of the previous section, from equation \eqref{betazz} we conclude that $W'=0$ at this point. On the other hand, regarding our discussion below the equation \eqref{betaz}, we should also insert $W''=0$ at this point. Imposing both conditions on \eqref{Bo4} or equivalently on its coefficients, we can read the superpotential of the bounce solution
\be 
W(\ph)=B_\pm\pm \frac{\sqrt{2}(d-1)}{15\k_1\sqrt{V_1}}(\ph-\ph_B)^\frac52+ \frac{d B_\pm}{90 \k_1 V_1}(\ph-\ph_B)^3+\cdots\,,
\ee
where $B_\pm$ is given in equation \eqref{bpm}.
According to this solution, the scalar field and scale factor near the critical point behave as
\begin{subequations}
\begin{align}
\phi(r) &=\phi_B+\frac12 V_1 (r-r_B)^2+\mathcal{O}(r-r_B)^3\,,\\
A(r) &=A_B-\frac{B_\pm}{2(d-1)}(r-r_B)-\frac{V_1^2}{720\k_1}(r-r_B)^6+\cdots\,.
\end{align}
\end{subequations} 
Therefore the direction of the RG flow is 
\be 
\frac{dW}{dr}=\frac{V_1^2(d-1)}{12}(r-r_B)^4+\cdots\,.
\ee
Again below and above the critical point $r=r_B$, the sign of $W'(r)$ does not change as we expect from a bounce point.
\subsubsection{\texorpdfstring{$W' \neq 0$ and $ W'' = 0$}{W'\unichar{"2260}0, W''=0}}
In this case, the shape of the bound near the critical point $\phi_B$ is given by 
$W_B$ in \eqref{bbb}. If we start from the following ansatz
\be 
W_B(\ph) =W_B+\mathcal{C}_1(\ph-\ph_B)+\mathcal{C}_z(\phi-\ph_B)^z+\cdots\,,
\ee
a perturbative analysis around the $\ph_B$ shows that $z=\frac52$. This can be confirmed by equation of motion for scalar field \eqref{phiz}. Then the coefficients are given by 
\begin{subequations} 
\begin{align}
\mathcal{C}_1&= \frac{16 (d-1)^4 V_0 + 4 (d-1)^3 d W_B^2 - d \k_2 W_B^4}{8 (d-1)^2 d \k_1 V_1 W_B}\,,\\
\mathcal{C}_z &= \pm \frac{2 (d-1)^2 -\mathcal{C}_1^2 d \k_1 }{15(d-1) \k_1 \sqrt{V_1}}\,.
\end{align}
\end{subequations}
Now if we compute the leading term of the scalar field we find that
\begin{subequations}
\begin{align}
\phi(r) &=\phi_B+ V_1 (r-r_B)^2+\mathcal{O}(r-r_B)^3\,.
\end{align}
\end{subequations}
However, for $V_1\neq 0$ equation \eqref{phiz} in addition to value of $z$ predicts the coefficient of the first term that should be $\frac12 V_1$. This behavior is happening in every bounce point that we have found up to now except this last one solution. In other words such a solution with $W' \neq 0$ and $ W'' = 0$ does not exist.
\section{Summary and conclusion}
In this work, in the context of gauge-gravity duality, we study the holographic RG flow of a CFT that is perturbed by a marginal/relevant operator. 
The bulk action contains the Einstein-dilaton gravity with an arbitrary scalar field potential together with the general quadratic curvature corrections \eqref{bulkaction}. 
Since the equations of motion are generally differential equations with more than two derivatives, see \eqref{EOMTrr} and \eqref{EOMTtt}, we may encounter the holographic RG equations that are not the first order either.
The study of these equations and their solutions may help us to understand the RG equations of the dual QFTs at their strong coupling regimes.

We do not impose any condition on the couplings, but we emphasize that our calculations are valid as far as we include all possible constraints or stability conditions on higher derivative theories. At least for small values of the couplings, we expect that one may avoid the ghost or tachyonic modes, which are integrated out and are beyond the QFT cut-off.

The equations of motion are classified by two combinations of the couplings, $\k_1$ and $\k_2$, in equation \eqref{k1k2}. 
As long as $\k_1$ is zero, we have a second-order differential equation of motion. The GB gravity is a particular case in this class. 

The sign of $\k_2$ plays an important role in the analysis of the critical points. For the negative values of $ \k_2$ in section \ref{31}, we observe similar behavior for holographic RG flow like the known EH case in  \cite{Kiritsis:2016kog}. The superpotential solutions near the local maxima (figure \ref{f1}) or minima (figure  \ref{f2}) of the potential are parameterizing  by some new coefficients as a function of $\k_2$. For example, the radii of $AdS$ solutions at the fixed points get modified according to \eqref{newparameters}. Moreover, the superpotential and its bound, $B_-$, in \eqref{bpm}, are shifted according to the value of this coupling. In section \ref{313}, similar to EH case, we observe the bounce solution \eqref{bkm} in this theory. For negative values of $\k_2$, one can show that the superpotential has the monotonic behavior \eqref{RG flow}, and so it can be used to construct a c-function in this theory. 

For positive values of $ \k_2 $, the situation is different. 
In addition, to the former lower bound on the superpotential, there is an upper bound, see figure \ref{f3}. 
According to the potential functionality, the values of the superpotential are more restricted, and they may confine in a specific region(s).

For both positive and negative values of $\k_2$, the curvature singularity removes if we restrict $W'$ to the finite values. However, for positive values of $\k_2$, a new condition should be added. The $W=W_E$ line \eqref{WEpm} is a boundary for the RG flows that divides the space of superpotential into two distinct upper and lower sub-spaces. We can find the superpotential near the local minima and maxima of the potential and the bounce solution on both the upper and lower bounds of the $W=W_E$ line. On this line, we have a curvature singularity. To avoid crossing this line and to have a smooth geometry, the RG flow should change its direction. 

For each positive value of $\k_2$, there is a lower bound for superpotential, which above that it has a monotonic behavior \eqref{RG flow}. 
If this situation holds, we expect the RG flow to reach asymptotically to the $W=W_E$ line at infinity. Otherwise, as we have shown in section  \eqref{WequalWE}, there are IR fixed points at the intersection of the upper and lower bounds, figure \ref{f4}, and therefore the flow ends on these points.

In regions below the $W=W_E$, the RG flow from a UV fixed point to an IR fixed point occurs such that $\tilde{L}_{UV}>\tilde{L}_{IR}$ where $\tilde{L}$'s are the radii of $AdS$ spaces at the fixed points.
In the upper regions $\tilde{L}_{UV}<\tilde{L}_{IR}$. 
The same behavior has been reported already in \cite{Ghodsi:2019xrx} via an ansatz for holographic RG flow. 
In this paper, the authors show that the a-anomaly, which is proportional to the number of degrees of freedom, decreases from UV to IR fixed points.  In other words, $a^*_{UV}>a^*_{IR}$, in both situations above. If we consider a monotonic superpotential, then we expect to have the same argument here.

In section \ref{4}, we consider the non-zero values of the $\k_1$. Although the equations of motion are more involved, we can find the perturbed solutions of the superpotential near the critical points of the theory. As we expect, since the equations of motion are fourth-order differential equations, there is more diversity in the solutions at fixed points that depend on the various parameters of the theory. The generic behavior of these solutions for fixed points near the local maxima of the potential, are shown in figures \ref{f5} to \ref{f10} and for local minima, in figures \ref{f11} to \ref{f16}. It is a general property that the local maxima of the potential are the UV fixed points, and the local minima could be either UV or IR fixed points.

According to the sign of $\k_2$, we may have two upper and lower bounds on the superpotential. 
However, the equations of the bounds are changed. We have a couple of constraints on superpotential and its derivative, \eqref{cons1}. 
The fixed singularity line $W=W_E$ in the previous case replaces by parabolic curves \eqref{SC}. These singular curves restrict the orientation of the superpotential. 
In other words, besides the value of the superpotential, the direction of the RG flow is controlled by the singular curves everywhere.

In the general theory, it is not clear that the superpotential remains a monotonic function.  Nevertheless, the equation \eqref{WPN} shows that it controls by the curvature of the domain wall solution. In other words, we can define a critical length at each point of the RG flow, and violation of a monotonous superpotential is related to the length scale in equation \eqref{Lc}.

\section*{Acknowledgment}
We would like to thanks E. Kiritsis for reading the manuscript and for his valuable comments. 
This work is supported by Ferdowsi University of Mashhad under the grant 2/54017 (1399/12/16).
\providecommand{\href}[2]{#2}\begingroup\raggedright

\endgroup

\begin{thebibliography}{99}
\bibitem{Wilson:1971bg}
K.~G.~Wilson,
``Renormalization group and critical phenomena. 1. Renormalization group and the Kadanoff scaling picture,''
\href{http://journals.aps.org/prb/abstract/10.1103/PhysRevB.4.3174}{Phys.\ Rev.\ B {\bf 4} (1971) 3174}.

\bibitem{Wilson:1971dh}
K.~G.~Wilson,
``Renormalization group and critical phenomena. 2. Phase space cell analysis of critical behavior,''
\href{http://journals.aps.org/prb/abstract/10.1103/PhysRevB.4.3184}{  Phys.\ Rev.\ B {\bf 4} (1971) 3184}.

\bibitem{Maldacena:1997re}
J.~M.~Maldacena,
``The Large N limit of superconformal field theories and supergravity,''
Adv. Theor. Math. Phys. \textbf{2} (1998), 231-252
\hre{hep-th}{9711200}.

\bibitem{Akhmedov:1998vf}
E.~T.~Akhmedov,
``A Remark on the AdS / CFT correspondence and the renormalization group flow,''
Phys. Lett. B \textbf{442} (1998), 152-158
\hre{hep-th}{9806217}.

\bibitem{Freedman:1999gp}
D.~Z.~Freedman, S.~S.~Gubser, K.~Pilch and N.~P.~Warner,
``Renormalization group flows from holography supersymmetry and a c theorem,''
Adv. Theor. Math. Phys. \textbf{3} (1999), 363-417
\hre{hep-th}{9904017}.

\bibitem{deBoer:1999tgo}
J.~de Boer, E.~P.~Verlinde and H.~L.~Verlinde,
``On the holographic renormalization group,''
JHEP \textbf{08} (2000), 003
\hre{hep-th}{9912012}.

\bibitem{Ceresole:2007wx}
A.~Ceresole and G.~Dall'Agata,
``Flow Equations for Non-BPS Extremal Black Holes,''
JHEP \textbf{03} (2007), 110
\hre{hep-th}{0702088}.

\bibitem{Kiritsis:2012ma}
E.~Kiritsis and V.~Niarchos,
``The holographic quantum effective potential at finite temperature and density,''
JHEP \textbf{08} (2012), 164
\hri{1205.6205}{[hep-th]}.

\bibitem{Bourdier:2013axa}
J.~Bourdier and E.~Kiritsis,
``Holographic RG flows and nearly-marginal operators,''
Class. Quant. Grav. \textbf{31} (2014), 035011
\hri{1310.0858}{[hep-th]}.

\bibitem{Kiritsis:2014kua}
E.~Kiritsis, W.~Li and F.~Nitti,
``Holographic RG flow and the Quantum Effective Action,''
Fortsch. Phys. \textbf{62} (2014), 389-454
\hri{1401.0888}{[hep-th]}.

\bibitem{Skenderis:1999mm}
K.~Skenderis and P.~K.~Townsend,
``Gravitational stability and renormalization group flow,''
Phys. Lett. B \textbf{468} (1999), 46-51
\hre{hep-th}{9909070}.

\bibitem{Girardello:1998pd}
L.~Girardello, M.~Petrini, M.~Porrati and A.~Zaffaroni,
``Novel local CFT and exact results on perturbations of N=4 superYang Mills from AdS dynamics,''
JHEP {\bf 9812} (1998) 022
\hre{hep-th}{9810126}.

\bibitem{Heemskerk:2010hk}
I.~Heemskerk and J.~Polchinski,
``Holographic and Wilsonian Renormalization Groups,''
JHEP \textbf{06} (2011), 031
\hri{1010.1264}{[hep-th]}.

\bibitem{Faulkner:2010jy}
T.~Faulkner, H.~Liu and M.~Rangamani,
``Integrating out geometry: Holographic Wilsonian RG and the membrane paradigm,''
JHEP \textbf{08} (2011), 051
\hri{1010.4036}{[hep-th]}.

\bibitem{Gursoy:2008za}
U.~Gursoy, E.~Kiritsis, L.~Mazzanti and F.~Nitti,
``Holography and Thermodynamics of 5D Dilaton-gravity,''
JHEP \textbf{05} (2009), 033
\hri{0812.0792}{[hep-th]}.

\bibitem{Gursoy:2007cb}
U.~Gursoy and E.~Kiritsis,
``Exploring improved holographic theories for QCD: Part I,''
JHEP \textbf{02} (2008), 032
\hri{0707.1324}{[hep-th]}.

\bibitem{Kiritsis:2016kog}
E.~Kiritsis, F.~Nitti and L.~Silva Pimenta,
``Exotic RG Flows from Holography,''
Fortsch.\ Phys.\  {\bf 65} (2017) no.2,  1600120
\hri{1611.05493}{[hep-th]}.

\bibitem{Ghosh:2017big}
J.~K.~Ghosh, E.~Kiritsis, F.~Nitti and L.~T.~Witkowski,
``Holographic RG flows on curved manifolds and quantum phase transitions,''
JHEP \textbf{05} (2018), 034
\hri{1711.08462}{[hep-th]}.

\bibitem{Bea:2018whf}
Y.~Bea and D.~Mateos,
``Heating up Exotic RG Flows with Holography,''
JHEP \textbf{08} (2018), 034
\hri{1805.01806}{[hep-th]}.

\bibitem{Gursoy:2018umf}
U.~G\"ursoy, E.~Kiritsis, F.~Nitti and L.~Silva Pimenta,
``Exotic holographic RG flows at finite temperature,''
JHEP \textbf{10} (2018), 173
\hri{1805.01769}{[hep-th]}.

\bibitem{Park:2019pzo}
C.~Park and J.~Hun Lee,
``Exotic RG flow of entanglement entropy,''
Phys. Rev. D \textbf{101} (2020) no.8, 086008
\hri{1910.05741}{hep-th]}.

\bibitem{Cremonini:2020rdx}
S.~Cremonini, L.~Li, K.~Ritchie and Y.~Tang,
``Constraining nonrelativistic RG flows with holography,''
Phys. Rev. D \textbf{103} (2021) no.4, 046006
\hri{2006.10780}{[hep-th]}.

\bibitem{Buchel:2009sk} 
  A.~Buchel, J.~Escobedo, R.~C.~Myers, M.~F.~Paulos, A.~Sinha and M.~Smolkin,
  ``Holographic GB gravity in arbitrary dimensions,''
  JHEP {\bf 1003}, 111 (2010)
\hri{0911.4257}{[hep-th]}.
  
\bibitem{Ghodsi:2018vhq}
S.~Qolibikloo and A.~Ghodsi,
``More on phase transition and R\'enyi entropy,''
Eur. Phys. J. C \textbf{79} (2019) no.5, 406
\hri{1811.04980}{[hep-th]}.

\bibitem{Brigante:2007nu}
M.~Brigante, H.~Liu, R.~C.~Myers, S.~Shenker and S.~Yaida,
``Viscosity Bound Violation in Higher Derivative Gravity,''
Phys. Rev. D \textbf{77} (2008), 126006
\hri{0712.0805}{[hep-th]}.


\bibitem{Brigante:2008gz}
M.~Brigante, H.~Liu, R.~C.~Myers, S.~Shenker and S.~Yaida,
``The Viscosity Bound and Causality Violation,''
Phys. Rev. Lett. \textbf{100} (2008), 191601
\hri{0802.3318}{[hep-th]}.

\bibitem{Ghodsi:2019xrx}
A.~Ghodsi and M.~Siahvoshan,
``A Holographic Study of the $a$-theorem and RG Flow in General Quadratic Curvature Gravity,''
Eur. Phys. J. C \textbf{79} (2019) no.10, 820
\hri{1907.03497}{[hep-th]}.

\bibitem{Ghodsi:2015gna}
A.~Ghodsi and M.~Moghadassi,
``Holographic entanglement entropy from minimal surfaces with/without extrinsic curvature,''
JHEP \textbf{02}, 037 (2016)
\hri{1508.02527}{[hep-th]}.

\bibitem{Ghodsi:2020qqb}
A.~Ghodsi, S.~Qolibikloo and S.~Karimi,
``Holographic complexity in general quadratic curvature theory of gravity,''
Eur. Phys. J. C \textbf{80}, no.10, 920 (2020)
\hri{2005.08989}{[hep-th]}.

\bibitem{Anastasiou:2021swo}
G.~Anastasiou, I.~J.~Araya, J.~Moreno, R.~Olea and D.~Rivera-Betancour,
``Renormalized holographic entanglement entropy for Quadratic Curvature Gravity,''
\hri{2102.11242}{[hep-th]}.

\bibitem{Boulware:1985wk}
D.~G.~Boulware and S.~Deser,
``String Generated Gravity Models,''
\href{http://journals.aps.org/prl/abstract/10.1103/PhysRevLett.55.2656}{Phys.\ Rev.\ Lett \textbf{55}, 2656 (1985)}.

\bibitem{Gullu:2009vy}
I.~Gullu and B.~Tekin,
``Massive Higher Derivative Gravity in D-dimensional Anti-de Sitter Spacetimes,''
Phys. Rev. D \textbf{80}, 064033 (2009)
\hri{0906.0102}{[hep-th]}.

\bibitem{Ghodsi:2017iee}
A.~Ghodsi and F.~Najafi,
``Ricci cubic gravity in d dimensions, gravitons and SAdS/Lifshitz black holes,''
Eur. Phys. J. C \textbf{77}, no.8, 559 (2017)
\hri{1702.06798}{[hep-th]}.


\end{thebibliography}
\end{document}